\documentclass[11pt]{article}

\usepackage[final]{acl}

\usepackage{times}
\usepackage{latexsym}

\usepackage[T1]{fontenc}

\usepackage[utf8]{inputenc}

\usepackage{microtype}

\usepackage{inconsolata}

\usepackage{graphicx}
\usepackage{dblfloatfix}

\usepackage{amssymb}
\usepackage{amsthm}
\usepackage{amsmath}
\usepackage{multirow}
\usepackage{booktabs}
\usepackage{makecell}
\usepackage{algorithm}
\usepackage{algpseudocode}
\usepackage{xcolor}
\usepackage{tabularx}
\usepackage{pifont}
\usepackage{array}
\usepackage{tcolorbox}
\usepackage{placeins}
\tcbuselibrary{breakable, skins}
\title{AgentMark: Utility-Preserving Behavioral Watermarking for Agents}

\author{
  \textbf{Kaibo Huang}$^{1}$\thanks{\ Equal contribution} \quad
  \textbf{Jin Tan}$^{1}$\footnotemark[1] \quad
  \textbf{Yukun Wei}$^{1}$ \quad
  \textbf{Wanling Li}$^{1}$ \quad
  \textbf{Zipei Zhang}$^{1}$ \\[0.2ex]
  \textbf{Hui Tian}$^{2}$ \quad
  \textbf{Zhongliang Yang}$^{1}$\thanks{\ Corresponding author} \quad
  \textbf{Linna Zhou}$^{1}$ \\[1ex]
  $^1$Beijing University of Posts and Telecommunications \\
  $^2$Huaqiao University \\
  \texttt{\{huangkaibo, tanjinanin, weiyukun, lwl, nebulazhang\}@bupt.edu.cn} \\
  \texttt{htian@hqu.edu.cn, \{yangzl, zhoulinna\}@bupt.edu.cn}
}

\begin{document}
\maketitle
\begin{abstract}
LLM-based agents are increasingly deployed to autonomously solve complex tasks, raising urgent needs for IP protection and regulatory provenance.
While content watermarking effectively attributes LLM-generated outputs, it fails to directly identify the high-level planning behaviors (e.g., tool and subgoal choices) that govern multi-step execution.
Critically, watermarking at the planning-behavior layer faces unique challenges: minor distributional deviations in decision-making can compound during long-term agent operation, degrading utility, and many agents operate as black boxes that are difficult to intervene in directly.
To bridge this gap, we propose AgentMark, a behavioral watermarking framework that embeds multi-bit identifiers into planning decisions while preserving utility.
 It operates by eliciting an explicit behavior distribution from the agent and applying distribution-preserving conditional sampling, enabling deployment under black-box APIs while remaining compatible with action-layer content watermarking.
Experiments across embodied, tool-use, and social environments demonstrate practical multi-bit capacity, robust recovery from partial logs, and utility preservation.
Code is available at \url{https://github.com/Tooooa/AgentMark}.
\end{abstract}

\section{Introduction}

\begin{figure}[t]
  \centering
  \includegraphics[width=\columnwidth]{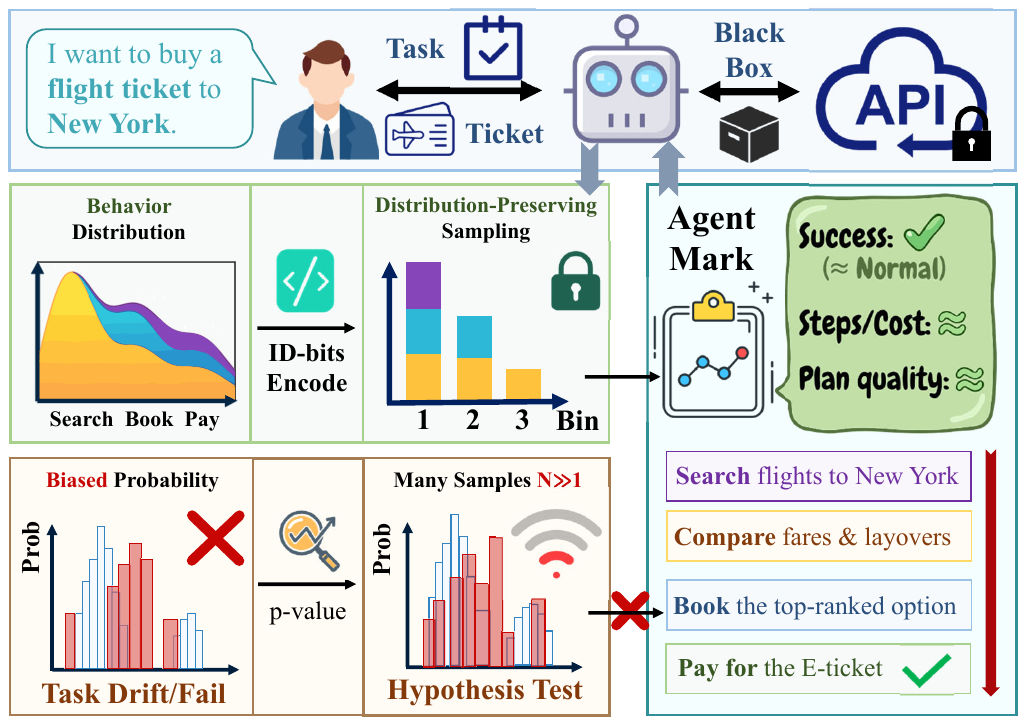}
    \caption{AgentMark embeds multi-bit provenance in planning behaviors via distribution-preserving sampling under black-box agent APIs, while preserving utility; bias-based probability watermarking can drift and harm task performance.}
  \label{fig:small}
\end{figure}

Recent advances in large language models (LLMs) have improved text generation and reasoning \cite{xu2025towards, achiam2023gpt, team2023gemini} and accelerated the shift from passive language interfaces to autonomous agents that can perceive context, plan, and execute actions \cite{shajarian2024survey, acharya2025agentic, yao2022react}. This autonomy and proactiveness are increasingly reflected in real-world applications, including GUI-based daily assistance \cite{liu2024autoglm}, tool-augmented multimodal agents for financial trading \cite{zhang2024multimodal}, and large-scale social agents operating in online ecosystems \cite{gao2024large}.

The increasing deployment of intelligent agents raises urgent concerns about traceability and accountability in regulated settings. In security-sensitive and high-impact domains, agents can be exploited for impersonation, automated disinformation, and large-scale manipulation, while enterprises face risks of misuse or unauthorized replication of proprietary agent systems \cite{gao2023s3, wang2025user, park2023generative, piao2025agentsociety, mou2024unveiling}. These challenges motivate provenance mechanisms that attribute agent activities and support auditing and enforcement.

To reason about provenance in agentic systems, we distinguish two levels of decision-making. \textbf{Planning behavior} refers to the high-level choice of \emph{what} to do next (e.g., selecting a tool or committing to a subgoal), whereas \textbf{execution action} specifies \emph{how} that choice is carried out (e.g., tool arguments or structured outputs). This behavior--action decomposition is widely used in embodied tasks and tool-use pipelines \cite{yao2022react, shajarian2024survey}. For simplicity, we use \textbf{behavior} and \textbf{action} as shorthand for \textbf{planning behavior} and \textbf{execution action} in the rest of the paper.

Content watermarking has shown strong practicality for attributing LLM-generated outputs, and recent systems such as SynthID-Text have been reported in production settings (e.g., Google Gemini) \cite{dathathri2024scalable}. However, agent provenance concerns not only the final content but also the agent’s behavioral decisions over time, which shape downstream impact and long-horizon outcomes, making behavioral watermarking a necessary complement to content watermarking. Directly extending existing content watermarking techniques to agent behaviors remains challenging. Training-time schemes that modify model weights are often infeasible because many agents rely on closed pre-trained APIs, and retraining general-purpose models is costly and hard to tailor to diverse environments \cite{lau2024waterfall, patil2023can, peng2023you}. Inference-time schemes that manipulate token-level sampling \cite{kirchenbauer2023watermark, dathathri2024scalable, guan2024codeip, hudecek-dusek-2023-large} also do not align well with high-level behaviors, since behaviors are not token-native and crucial semantics can be lost when a planning decision is compiled into structured executions (e.g., tool calls and arguments).  For example, an LLM output such as ``Alice bookmarked a post with the tag \texttt{\#TravelInspiration}'' may be reduced to discrete actions like ``bookmarking'' and ``tagging,'' stripping tokens that are useful for provenance, which makes watermark extraction and verification more difficult. Moreover, just watermarking behaviors by directly biasing the behavior probabilities \cite{huang2025agent} can introduce compounding distribution shifts over long horizons, leading to task drift or failure. These gaps motivate robust watermarking methods that operate at the behavior level while preserving task utility.

As illustrated in Figure~\ref{fig:small}, we address this gap by introducing AgentMark, a behavioral watermarking framework that embeds multi-bit identifiers into an agent's planning process with utility preservation. Viewing planning as sampling from a time-varying behavior distribution, where the agent would otherwise make an implicit behavior choice, AgentMark first elicits an explicit probability list over candidate behaviors and then embeds a watermark by distribution-preserving sampling on this elicited distribution, keeping the induced behavior distribution unchanged. To withstand agent-specific failures such as step erasure and truncation (e.g., platform filtering and logging loss), we use context-reproducible randomness and erasure-resilient coding, enabling recovery from partial trajectories. Our contributions are as follows:
\begin{itemize}
\item We propose a planning-level framework for agent behavioral watermarking, where we elicit behavior distributions and encode multi-bit IDs via distribution-preserving sampling, without changing model weights or token-level sampling.
\item We implement this framework as AgentMark-F, using context-reproducible randomness and erasure-resilient decoding to recover IDs from partial trajectories.
\item We evaluate AgentMark in embodied, tool-use, and social environments, showing practical multi-bit capacity with utility preservation, robustness to erasure or truncation and semantic-preserving observation rewriting, and compatibility with action-layer content watermarking.
\end{itemize}

\section{Related Work}

\subsection{Content Watermarking for LLMs}
Content watermarking attributes LLM-generated text by embedding detectable signals during generation, supported by standardized toolchains such as MarkLLM~\citep{pan2024markllm}.
Prior work spans model-level watermarking \citep[e.g.,][]{zhu2024reliable} and decoding-time schemes, including statistical logit/sampling signals \citep[e.g.,][]{kirchenbauer2023watermark}, distribution-preserving designs \citep[e.g.,][]{dathathri2024scalable, chen2025improved}, semantic robustness to rewriting/translation \citep[e.g.,][]{liu2024semanticinvariantrobustwatermark, he2024watermarkssurvivetranslationcrosslingual, hou2024semstamp}, and multi-bit identification \citep[e.g.,][]{lee2024wrote}.
While highly effective for attributing free-form text, content watermarking operates on the action/content layer and does not provide direct provenance for planning-time behaviors in agent pipelines, motivating complementary watermarking mechanisms at the behavior layer.

\subsection{Agent Security}
Autonomous agents deployed in open ecosystems introduce security and governance risks, including impersonation, automated manipulation, and tool-enabled abuse \cite{he2025security, deng2025ai}, as well as unauthorized replication of proprietary agent systems \cite{triedman2025multi}. The emerging Internet of Agents (IoA) \cite{wang2025internet} further raises a potential risk of covert harmful communication across agent interactions \cite{huang2025whispering}.
But, strengthening agent safety often incurs a non-trivial \emph{security tax} in capability, usability, or cost \cite{yang2024towards}, complicating real-world adoption.
These challenges motivate provenance mechanisms that enable reliable attribution and auditing while preserving agent utility.

\section{Problem Formulation}

\subsection{Preliminaries and Notation}
We consider an agent interacting with an environment over a trajectory of length $T$.
At each time step $t \in \{1,\ldots,T\}$, the agent observes $o_t$ and is given a finite set of admissible planning behaviors $\mathcal{B}_t$.
The agent selects a planning behavior $b_t \in \mathcal{B}_t$ and then executes an \emph{execution action} $a_t$ that specifies how $b_t$ is carried out.

We assume the agent induces a latent planning-time behavior distribution $P_t^\star$ over $\mathcal{B}_t$ and that the baseline makes a behavior choice accordingly.
In our setting, we elicit an explicit estimate $P_t$ of $P_t^\star$ and define distribution preservation with respect to $P_t$.
Our watermarking operates at the \emph{behavior} level and does not directly manipulate the execution action $a_t$.
To avoid shifting the planning policy, we require per-step distribution preservation:
\begin{equation}
\hat b_t \sim P_t \quad \text{for all } t \in \{1,\ldots,T\}.
\end{equation}

\subsection{Planning-Time Behavior Channel and Distribution-Preserving Coding}
\label{sec:behavior-channel}
We model the agent's planning stage as sampling from a time-varying discrete channel. At each time step $t$, the channel is characterized by an implicit behavior distribution $P_t^\star$ over the behavior set $\mathcal{B}_t$. The unwatermarked agent makes an implicit behavior choice $b_t \sim P_t^\star$. In our setting, we elicit an explicit estimate of $P_t^\star$ as a probability list $P_t$ over candidate behaviors. Our distribution-preserving guarantee is defined with respect to the elicited distribution $P_t$. To embed a provenance payload $m \in \{0,1\}^L$, a watermarked agent replaces random sampling with a distribution-preserving encoder and outputs
\begin{equation}
\hat b_t \leftarrow \mathsf{Enc}(P_t, r_t) \in \mathcal{B}_t,
\end{equation}
where $r_t \in (0,1]$ denotes shared sampling randomness. We assume the embedder and verifier share a secret key $K_{\mathrm{sh}}$ so that the verifier can reproduce the same per-step randomness from logged context when decoding.

The encoder consumes bits from $m$ across steps depending on $P_t$.
Given the logged behavior sequence and the corresponding channel information, the verifier recovers the payload via
\begin{equation}
\hat m \leftarrow \mathsf{Dec}\big(\{(\hat b_t, P_t)\}_{t=1}^T\big).
\end{equation}

\begin{figure*}[t]
  \centering
  \includegraphics[width=\textwidth]{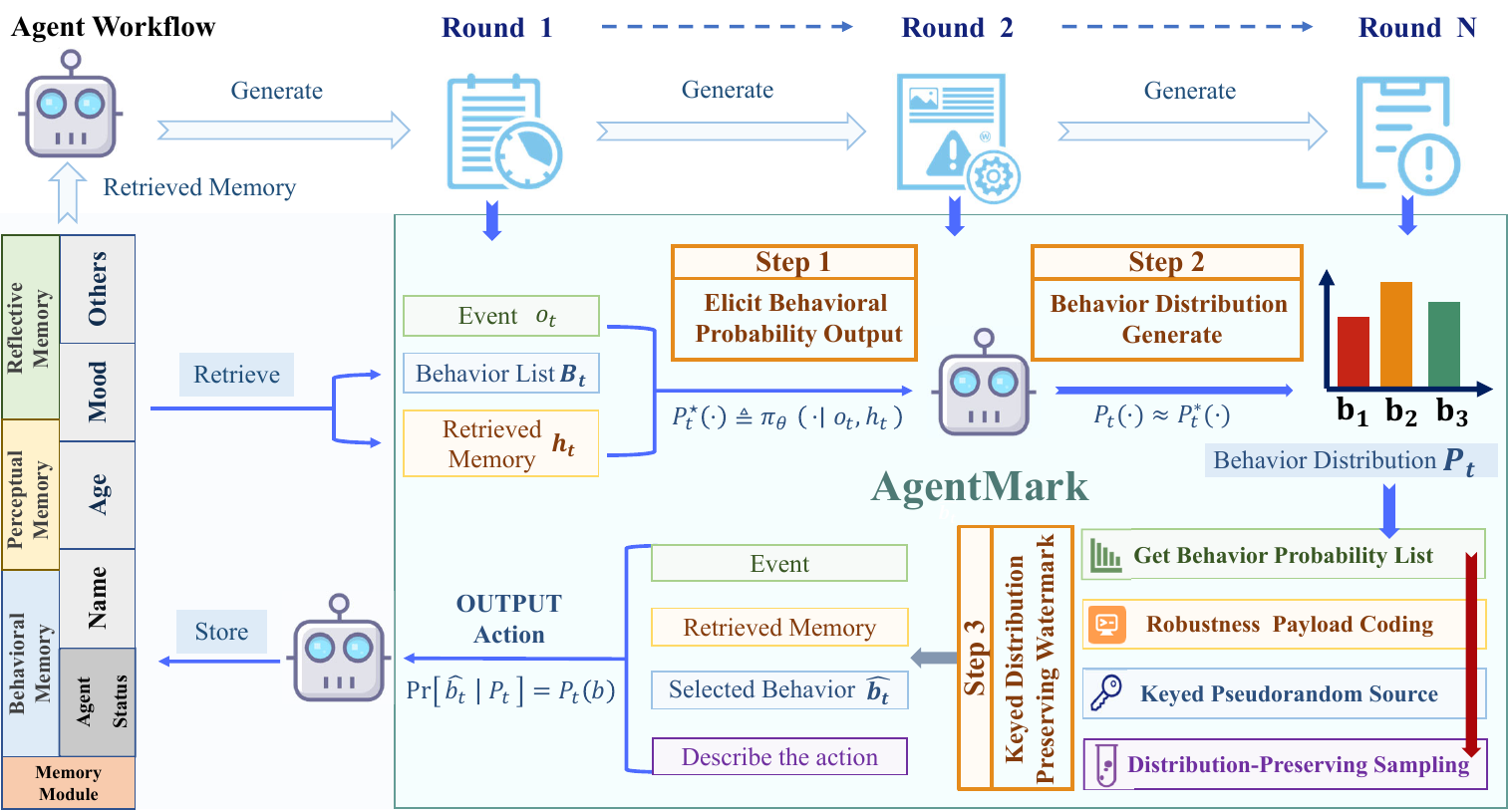}
  \caption{AgentMark overview. At each round, the agent would otherwise make an implicit planning-behavior choice according to a latent policy $P_t^\star$ over a finite behavior set $\mathcal{B}_t$. AgentMark makes this decision process auditable by eliciting an explicit probability list $P_t(\cdot)\approx P_t^\star(\cdot)$ over $\mathcal{B}_t$, and then applies distribution-preserving watermark sampling on $P_t$ to select a planning behavior $\hat b_t$ while keeping its marginal distribution matched to $P_t$. The execution action is generated conditioned on $\hat b_t$ and executed in the environment.}
  \label{fig:overview}
\end{figure*}

\subsection{Threat Model}
We consider provenance verification in agentic workflows, where a user specifies a high-level goal, and the agent autonomously executes a multi-step plan with limited human intervention \cite{wiesinger2024agents}. In such pipelines, the verifier may not receive a complete record of planning-time decisions. Intermediate steps can be missing due to incomplete logging, execution failures, or platform-side deletion, leading to step erasure and trajectory truncation.

We model this as an incomplete observation of a length-$T$ trajectory, where $\mathcal{I} \subseteq \{1,\ldots,T\}$ denotes the set of observed time indices.
The verifier observes only $\{(\hat b_t, P_t)\}_{t\in\mathcal{I}}$, while steps in $\{1,\ldots,T\}\setminus\mathcal{I}$ are missing. We parameterize missingness by an erasure rate $\rho\in[0,1)$, with $|\mathcal{I}| \approx (1-\rho)T$ (or $|\mathcal{I}| \ge (1-\rho)T$ in a worst-case view).
Truncation corresponds to observing only a prefix, i.e., $\mathcal{I}=\{1,\ldots,\tau\}$ for some $\tau<T$ (more generally, a contiguous segment may be observed).

Post-hoc modifications to \emph{execution actions} (e.g., editing tool arguments) may occur in practice. Since our watermark is embedded in behaviors rather than action surface forms, we treat the main effect of such interventions as missing or shortened planning-time records, and we formalize robustness under erasure and truncation accordingly.

\subsection{Objectives}
\label{sec:objectives}
We formalize two objectives: utility preservation and robust decodability.
Our design enforces per-step distribution preservation as a constraint to support utility preservation.

\paragraph{Utility preservation.}
We require that watermarking does not degrade task execution quality or efficiency, measured by task success and trajectory length:
\begin{equation}
\left\{
\begin{aligned}
\Big|\mathbb{E}[\mathrm{Succ}(\hat\tau)]-\mathbb{E}[\mathrm{Succ}(\tau)]\Big| &\le \varepsilon_{\mathrm{succ}}\\
\Big|\mathbb{E}[\mathrm{Len}(\hat\tau)]-\mathbb{E}[\mathrm{Len}(\tau)]\Big| &\le \varepsilon_{\mathrm{len}}
\end{aligned},
\right.
\end{equation}
where $\tau$ and $\hat\tau$ denote the baseline and watermarked trajectories induced by the baseline agent's implicit behavior choices $b_t$ and the watermarked choices $\hat b_t$, respectively; $\mathrm{Succ}(\cdot)\in\{0,1\}$ indicates whether the task is completed successfully, and $\mathrm{Len}(\cdot)$ is the number of decision steps (a proxy for time or compute overhead).

\paragraph{Robust decodability.}
Under step erasure or truncation, the verifier observes indices $\mathcal{I}\subseteq\{1,\ldots,T\}$ and receives $\{(\hat b_t,P_t)\}_{t\in\mathcal{I}}$.
For a target erasure rate $\rho$, we require successful recovery from partial trajectories whenever $|\mathcal{I}| \ge (1-\rho)T$:
\begin{equation}
\Pr\!\left[\mathsf{Dec}\big(\{(\hat b_t,P_t)\}_{t\in\mathcal{I}}\big)=m\right]\ge 1-\delta,
\end{equation}
where $m\in\{0,1\}^L$ is the payload and $\delta$ is the allowed decoding failure probability.

\section{AgentMark}

\subsection{Overview}
Figure~\ref{fig:overview} illustrates AgentMark in an agent workflow.
A key design requirement in agentic settings is that watermarking should not reduce task performance; accordingly, AgentMark operates on the planning layer and preserves the elicited planning distribution to avoid compounding errors over long horizons.
At each round, an LLM agent would otherwise make an implicit planning-behavior choice according to a latent policy $\pi_\theta$  over a finite candidate set $\mathcal{B}_t$.
AgentMark makes this decision process auditable by eliciting an explicit probability list $P_t$ over $\mathcal{B}_t$, and then watermarking only through how the planning behavior is sampled from $P_t$. Formally, we denote the LLM-induced implicit planning policy by
\begin{equation}
P_t^\star(\cdot)\triangleq \pi_\theta(\cdot\mid o_t,h_t)\in \Delta(\mathcal{B}_t),
\end{equation}
which is typically latent in standard agent implementations.

AgentMark makes this policy auditable by eliciting an explicit estimate as a probability list over the same candidate set:
\begin{equation}
P_t(\cdot)\approx P_t^\star(\cdot)\in \Delta(\mathcal{B}_t).
\end{equation}
This explicit access enables distribution-preserving sampling and verification.

\paragraph{Behavior--action separation.}
AgentMark outputs a watermarked planning behavior $\hat b_t$, after which the execution action is generated conditioned on $\hat b_t$ and executed in the environment.
Rather than manipulating action, AgentMark intervenes only in the planning-stage sampling procedure by using distribution-preserving sampling on the elicited behavior distribution.
In particular, it preserves the planning-time behavior distribution, i.e., the marginal of $\hat b_t$ matches the original elicited distribution $P_t$, so watermarking does not materially shift the planning distribution used at inference time or alter the agent's decision logic.

\paragraph{Robustness interface.}
To improve robustness under missing steps, AgentMark optionally applies payload coding before sampling and supports erasure-resilient recovery from partial trajectories.

\subsection{Distribution-Preserving Behavioral Watermark}
At each step $t$, given an explicit behavior distribution $P_t$ over $\mathcal{B}_t$ and a payload bitstream $M$, the encoder selects a watermarked behavior while preserving the marginal distribution:
\begin{equation}
 \left\{\begin{matrix}
\hat b_t \leftarrow \mathsf{Enc}(P_t, M, r_t) \in \mathcal{B}_t \\
\Pr[\hat b_t=b] = P_t(b), \forall b\in\mathcal{B}_t
\end{matrix}\right.,
\end{equation}
where $r_t$ denotes shared sampling randomness.
The decoder inverts the same sampling process to extract the embedded bits from an observed behavior sequence.

\paragraph{Keyed pseudorandomness and per-step independence.}
Following the setting in Section~\ref{sec:behavior-channel}, we derive per-step randomness from a shared secret $K_{\mathrm{sh}}$ and step context $Context_t$ to synchronize encoding and decoding, where $Context_t$ summarizes the information shared by both sides at time $t$ (e.g., the step index $t$, the observation $o_t$, and the agent history/context $h_t$).
We derive
\begin{equation}
K_t \leftarrow H(K_{\mathrm{sh}} \,\|\, Context_t),
\end{equation}
and instantiate a pseudorandom generator seeded by $K_t$ to produce the randomness stream $r_t$ used by the sampler.

\subsection{AgentMark-F: A Concrete Instantiation of AgentMark.}
AgentMark is a general paradigm that embeds provenance bits into planning behaviors via distribution-preserving conditional sampling on an explicit behavior distribution $P_t$.
Several distribution-preserving sampling schemes for conditional distributions have been developed (e.g., Meteor~\citealp{kaptchuk2021meteor} and Discop~\citealp{ding2023discop}).
\citet{liao2025framework} further proposes FDPSS, a modular framework with a security proof.
Following this line, we instantiate AgentMark with an FDPSS-style construction, termed \textbf{AgentMark-F}. Concretely, AgentMark-F transforms $P_t$ into a mixture of uniform bins via probability recombination, samples a bin using keyed pseudorandomness, and applies cyclic-shift uniform encoding within the selected bin to embed a variable-length bitstring.

\paragraph{Differential recombination and distribution preservation.}
At each step $t$, let $\mathcal{B}_t=\{b_{t,1},\ldots,b_{t,n}\}$ be the candidate behavior set.
We first form a canonical ordering of $\mathcal{B}_t$ by sorting behaviors by their probabilities in non-increasing order, yielding $p_1 \ge \cdots \ge p_n$ where $p_i=P_t(b_{t,i})$.
Differential-based recombination decomposes the sorted probability sequence via first-order differences, defining slice heights $d_k \triangleq p_k - p_{k+1}$ with $p_{n+1}=0$ (for the full proof and details of distribution preservation, see Appendix~\ref{appendix:recombination}).
Differential recombination constructs a mixture of uniform bins in which bin $k$ contains the top-$k$ behaviors and has mixture weight $q_k = k \cdot d_k$.
Given index $K\sim\mathrm{Cat}(q_1,\ldots,q_n)$, the encoder samples uniformly within bin $K$ while embedding bits. This preserves the marginal: for any $i$,
\begin{equation}
\begin{aligned}
\Pr[\hat b_t=b_{t,i}]
&= \sum_{k=i}^{n} q_k \cdot \frac{1}{k} \\
&= \sum_{k=i}^{n} (p_k - p_{k+1}) \\
&= p_i .
\end{aligned}
\end{equation}

\paragraph{Encoder and Decoder.}
AgentMark-F uses keyed pseudorandomness to synchronize bin sampling and uniform encoding between the encoder and decoder.
At step $t$, both sides derive a per-step seed from the shared secret and context, sample the same bin according to $\{q_k\}$, and apply CyclicShift within the bin to embed/extract a variable-length bitstring.
For a bin $T$ of size $n=|T|$, CyclicShift embeds $c_t\in\{\lfloor \log_2 n\rfloor,\ \lfloor \log_2 n\rfloor+1\}$ bits, with near-optimal expected capacity $\mathbb{E}[c_t\mid |T|=n]\in[\log_2 n-0.0861,\log_2 n]$.
Across $t=1,\ldots,T$, the encoder repeats Algorithm~\ref{alg:dp-encode} with a payload pointer $\ell$, producing substrings $\{s_t\}$ whose lengths depend on $|T|$.
Throughout, we embed information while preserving the marginal behavior:
\begin{equation}
\Pr[\hat b_t=b \mid P_t]  = P_t(b), \quad \forall b\in\mathcal{B}_t.
\end{equation}

Details are in Appendix~\ref{appendix:cyclicshift} (example: Appendix~\ref{appendix:example}); the full distribution-preservation proof is deferred to Appendix~\ref{appendix:Security_Proof}.

\begin{algorithm}[t]
\caption{AgentMark-F (one step): \textsc{Encode}}
\label{alg:dp-encode}
\begin{algorithmic}[1]
\Require $P_t$ over $\mathcal{B}_t$, step context $Context_t$, shared secret $K_{\mathrm{sh}}$, payload bitstream $M$, pointer $\ell$
\Ensure $\hat b_t$, embedded bits $s_t$, updated pointer $\ell$
\State $K_t \gets H(K_{\mathrm{sh}} \,\|\, Context_t)$; $\mathsf{PRG}\gets \mathsf{PRG}(K_t)$
\State $\{(T_k,q_k)\}_{k=1}^{n} \gets \mathsf{DiffRecombine}(P_t)$
\State $K \gets \mathsf{SampleCat}(\{q_k\}_{k=1}^{n}, \mathsf{PRG})$; $T \gets T_K$
\State $(j,s_t) \gets \mathsf{CyclicShiftEnc}(M[\ell:], |T|, \mathsf{PRG})$
\State $\hat b_t \gets T[j]$; $\ell \gets \ell + |s_t|$
\State \Return $\hat b_t, s_t, \ell$
\end{algorithmic}
\end{algorithm}

\begin{algorithm}[t]
\caption{AgentMark-F (one step): \textsc{Decode}}
\label{alg:dp-decode}
\begin{algorithmic}[1]
\Require $\hat b_t$, $P_t$ over $\mathcal{B}_t$, step context $Context_t$, shared secret $K_{\mathrm{sh}}$
\Ensure extracted bitstring $s_t$ (possibly empty)
\State $K_t \gets H(K_{\mathrm{sh}} \,\|\, Context_t)$; $\mathsf{PRG}\gets \mathsf{PRG}(K_t)$
\State $\{(T_k,q_k)\}_{k=1}^{n} \gets \mathsf{DiffRecombine}(P_t)$
\State $K \gets \mathsf{SampleCat}(\{q_k\}_{k=1}^{n}, \mathsf{PRG})$; $T \gets T_K$
\State $j \gets \mathsf{IndexOf}(\hat b_t \text{ in } T)$
\State $s_t \gets \mathsf{CyclicShiftDec}(j, |T|, \mathsf{PRG})$
\State \Return $s_t$
\end{algorithmic}
\end{algorithm}

\subsection{Erasure-Resilient Coding and Decoding}
\paragraph{Variable-capacity bitstream.}
At step $t$, the sampler outputs $s_t\in\{0,1\}^{c_t}$ with
\begin{equation}
c_t \triangleq |s_t| \in \Big\{\lfloor \log_2 |T| \rfloor,\ \lfloor \log_2 |T| \rfloor + 1\Big\},
\end{equation}
where $|T|$ is the selected bin size (Appendix~\ref{appendix:cyclicshift}).
Under step erasure/truncation, the verifier observes indices $\mathcal{I}\subseteq\{1,\ldots,T\}$ and receives a total of $R \triangleq \sum_{t\in\mathcal{I}} c_t$
embedded bits.

\paragraph{RLNC \cite{ho2006random} over $\mathbb{F}_2$.}
Let the provenance payload be $m\in\mathbb{F}_2^{L}$.
At step $t$, we derive
\begin{equation}
K_t \leftarrow H(K_{\mathrm{sh}} \,\|\, Context_t),
\end{equation}
and deterministically generate coefficient vectors
\begin{equation}
a_{t,j} \leftarrow \mathsf{PRG}(K_t,j)\in\mathbb{F}_2^{L}, j=1,\ldots,c_t.
\end{equation}
We pseudorandomize the RLNC coefficients and embedded bits using $K_t$; see Appendix~\ref{appendix:Security_Proof} for the security argument and Appendix~\ref{appendix:rlnc} for additional mathematical details.
Each embedded bit is a linear equation
\begin{equation}
y_{t,j} \triangleq \langle a_{t,j}, m\rangle \in \mathbb{F}_2,
\end{equation}
and $s_t=(y_{t,1},\ldots,y_{t,c_t})$.

\paragraph{Decoding under erasure/truncation.}
From $\mathcal{I}$, the verifier reconstructs $\{a_{t,j}\}$ and forms
\begin{equation}
A_{\mathcal{I}}\, m = y_{\mathcal{I}} \quad \text{over } \mathbb{F}_2,
\end{equation}
where $A_{\mathcal{I}}\in\mathbb{F}_2^{R\times L}$ stacks $\{a_{t,j}\}_{t\in\mathcal{I},\,1\le j\le c_t}$ as rows and $y_{\mathcal{I}}\in\mathbb{F}_2^{R}$ stacks the corresponding $\{y_{t,j}\}$.
Unique recovery holds iff $\mathrm{rank}(A_{\mathcal{I}})=L$. When rows are pseudorandom in $\mathbb{F}_2^{L}$, for $R=r\ge L$ we have the rank bound.
\begin{equation}
\Pr[\mathrm{rank}(A_{\mathcal{I}})=L \mid R=r] \ge 1-2^{-(r-L)}.
\label{eq:rlnc-rank-bound}
\end{equation}
Therefore, the decoding success probability admits the lower bound
\begin{equation}
\Pr[\hat m=m] \ge \sum_{r=L}^{\infty} \Pr[R=r]\cdot \big(1-2^{-(r-L)}\big),
\label{eq:rlnc-success-lb}
\end{equation}
which depends on the observed set $\mathcal{I}$ only through $R=\sum_{t\in\mathcal{I}}c_t$, matching our variable capacity.

\begin{table*}[t]
\centering
\small
\setlength{\tabcolsep}{3.0pt}
\renewcommand{\arraystretch}{1.02}
\resizebox{\textwidth}{!}{
\begin{tabular}{lc
                ccc ccc
                cccc}
\toprule
\multirow{2}{*}{Setting} & \multirow{2}{*}{Task} &
\multicolumn{3}{c}{SR (\%) $\uparrow$} &
\multicolumn{3}{c}{Steps $\downarrow$} &
\multicolumn{4}{c}{Watermark}  \\
\cmidrule(lr){3-5}\cmidrule(lr){6-8}\cmidrule(lr){9-12}
& &
Base & RG & Ours &
Base & RG & Ours &
\makecell[c]{bps $\uparrow$} & \makecell[c]{bpt $\uparrow$} &
\makecell[c]{$\Delta$s/step $\downarrow$} &
\makecell[c]{$\Delta$Tok/step (\%) $\downarrow$} \\
\midrule

\multirow{7}{*}{\makecell{ALFWorld\\ID}}
& A1 & 97.1$\pm$16.7 & 96.2$\pm$19.2 & 97.1$\pm$16.7 & 12.8$\pm$10.8 & 13.8$\pm$10.6 & 11.5$\pm$9.6  & 1.19 & 14.2 & $+0.44$ & $-0.20\%$ \\
& A2 & 98.9$\pm$10.3 & 100.0$\pm$0.0 & 100.0$\pm$0.0 & 10.2$\pm$7.0  & 11.0$\pm$6.8  & 11.0$\pm$7.7  & 1.06 & 14.0 & $+0.29$ & $+0.25\%$ \\
& A3 & 87.3$\pm$33.3 & 94.4$\pm$22.9 & 87.3$\pm$33.3 & 20.1$\pm$14.8 & 16.4$\pm$11.4 & 18.6$\pm$15.0 & 1.23 & 30.7 & $+0.05$ & $+0.12\%$ \\
& A4 & 89.7$\pm$30.3 & 82.0$\pm$38.4 & 91.0$\pm$28.6 & 18.2$\pm$13.5 & 20.5$\pm$15.4 & 17.2$\pm$13.2 & 1.17 & 26.4 & $-0.14$ & $+0.42\%$ \\
& A5 & 94.4$\pm$22.9 & 93.8$\pm$24.2 & 88.9$\pm$31.4 & 17.2$\pm$11.9 & 16.2$\pm$9.9  & 19.4$\pm$13.4 & 1.37 & 34.3 & $+0.26$ & $-0.08\%$ \\
& A6 & 78.5$\pm$41.1 & 60.6$\pm$48.9 & 77.6$\pm$41.7 & 31.9$\pm$14.8 & 39.5$\pm$12.9 & 31.3$\pm$15.6 & 1.28 & 37.1 & $-0.13$ & $+0.19\%$ \\
& \textbf{Avg.}
& \textbf{89.5$\pm$30.6} & \textbf{78.8($\downarrow 10.7$)} & \textbf{89.3($\downarrow 0.2$)}
& \textbf{19.7$\pm$14.9} & \textbf{26.1($\uparrow 6.4$)} & \textbf{19.4($\downarrow 0.3$)}
& \textbf{1.19} & \textbf{25.5} & $\textbf{+0.10}$ & $\textbf{+0.20\%}$ \\
\midrule

\multirow{7}{*}{\makecell{ALFWorld\\OOD}}
& A1 & 99.4$\pm$7.6  & 99.4$\pm$7.6  & 99.4$\pm$7.6  & 14.5$\pm$6.6  & 11.4$\pm$8.1  & 11.0$\pm$6.9  & 1.38 & 30.3 & $+0.07$ & $-0.76\%$ \\
& A2 & 100.0$\pm$0.0 & 93.8$\pm$24.2 & 97.9$\pm$14.3 & 12.2$\pm$6.9  & 15.1$\pm$12.1 & 13.6$\pm$9.4  & 1.39 & 25.5 & $+0.15$ & $-0.22\%$ \\
& A3 & 91.9$\pm$27.2 & 90.3$\pm$29.6 & 96.8$\pm$17.7 & 18.5$\pm$12.7 & 17.0$\pm$12.7 & 15.2$\pm$10.0 & 1.29 & 24.5 & $-0.18$ & $+0.09\%$ \\
& A4 & 97.6$\pm$15.2 & 90.5$\pm$29.4 & 97.6$\pm$15.2 & 16.2$\pm$10.0 & 17.1$\pm$12.3 & 14.3$\pm$9.1  & 1.18 & 22.7 & $-1.01$ & $-0.11\%$ \\
& A5 & 91.3$\pm$28.2 & 89.1$\pm$31.1 & 91.3$\pm$28.2 & 16.4$\pm$12.6 & 19.3$\pm$14.7 & 17.1$\pm$12.5 & 1.44 & 33.4 & $+0.02$ & $-0.37\%$ \\
& A6 & 94.1$\pm$23.5 & 91.2$\pm$28.4 & 97.1$\pm$16.9 & 22.4$\pm$11.1 & 25.4$\pm$13.3 & 24.1$\pm$11.8 & 1.41 & 38.2 & $-0.18$ & $-0.88\%$ \\
& \textbf{Avg.}
& \textbf{96.8$\pm$17.7} & \textbf{94.5($\downarrow 2.3$)} & \textbf{97.5($\uparrow 0.7$)}
& \textbf{15.9$\pm$9.7}  & \textbf{15.4($\downarrow 0.5$)} & \textbf{14.1($\downarrow 1.8$)}
& \textbf{1.34} & \textbf{28.4} & $\textbf{-0.18}$ & $\textbf{-0.30\%}$ \\
\midrule

\multirow{7}{*}{ToolBench}
& T1 & 61.7$\pm$2.4 & 60.0$\pm$4.1 & 60.0$\pm$7.1 & 5.2$\pm$5.1 & 5.4$\pm$4.9 & 5.6$\pm$4.7 & 0.51 & 5.28 & $-1.67$ & $+9.64\%$ \\
& T2 & 58.3$\pm$6.2 & 61.7$\pm$10.3 & 61.7$\pm$2.4 & 7.1$\pm$5.3 & 8.7$\pm$5.4 & 8.6$\pm$4.8 & 0.48 & 4.89 & $-1.41$ & $-16.96\%$ \\
& T3 & 66.7$\pm$9.4 & 60.0$\pm$4.1 & 60.0$\pm$4.1 & 4.7$\pm$4.0 & 4.8$\pm$4.1 & 4.8$\pm$4.6 & 0.46 & 4.62 & $-2.40$ & $-3.29\%$ \\
& T4 & 73.3$\pm$6.2 & 66.7$\pm$4.7 & 71.7$\pm$4.7 & 5.7$\pm$3.8 & 6.3$\pm$4.4 & 6.5$\pm$4.3 & 0.49 & 5.10 & $-1.54$ & $+10.97\%$ \\
& T5 & 49.3$\pm$7.4 & 49.3$\pm$4.2 & 50.0$\pm$3.5 & 8.1$\pm$5.6 & 7.7$\pm$5.4 & 7.8$\pm$5.6 & 0.48 & 4.85 & $-0.83$ & $+5.24\%$ \\
& T6 & 50.0$\pm$4.1 & 53.3$\pm$2.4 & 55.0$\pm$7.1 & 9.2$\pm$5.4 & 9.3$\pm$5.7 & 9.6$\pm$6.1 & 0.49 & 4.90 & $+0.24$ & $-14.77\%$ \\
& \textbf{Avg.}
& \textbf{59.9$\pm$5.8} & \textbf{58.5($\downarrow 1.4$)} & \textbf{59.7($\downarrow 0.2$)}
& \textbf{6.7$\pm$4.9}  & \textbf{7.0($\uparrow 0.3$)}  & \textbf{7.2($\uparrow 0.5$)}
& \textbf{0.49} & \textbf{4.93} & $\textbf{-1.27}$ & $\textbf{-6.25\%}$ \\
\bottomrule
\end{tabular}
}
\caption{ALFWorld and ToolBench results. We report success rate (SR; fraction of tasks solved) and Steps (average decision steps over successful episodes). For Ours, we report watermark capacity bps/bpt (bits/step and bits/task), per-step latency difference $\Delta$s/step (Ours--Base), and token/step change rate $\Delta$Tok/step $=(\mathrm{Tok/step}_{\mathrm{Ours}}-\mathrm{Tok/step}_{\mathrm{Base}})/\mathrm{Tok/step}_{\mathrm{Base}}$. ALFWorld A1--A6 map to Look At Obj In Light, Pick And Place Simple, Pick Clean Then Place, Pick Cool Then Place, Pick Heat Then Place, and Pick Two Obj And Place; ToolBench T1--T6 map to single-tool selection, single-tool instruction-following, single-tool within-category selection, multi-tool within-category selection, multi-tool instruction-following, and complex multi-tool instruction-following. Appendix~\ref{sec:capacity_entropy} provides a complementary step-level capacity--entropy analysis on archived elicitation logs.}
\label{tab:main_results}
\end{table*}

\section{Experiment}

\subsection{Experimental Setup}

\paragraph{Environments and tasks.}
We evaluate AgentMark in three deployment-relevant settings that cover embodied planning in ALFWorld~\cite{shridhar2020alfworld} (ID/OOD splits; OOD is a cross-distribution test set), tool-use decision making on ToolBench~\cite{qin2023toolllm} (six test subsets spanning single- and multi-tool regimes), and social simulation in OASIS~\cite{yang2024oasis} on both Twitter-like and Reddit-like platforms; further dataset details are provided in Appendix~\ref{appendix:datasets}.

\paragraph{Compared methods.}
\textbf{Baseline} follows the ReAct agent loop and directly outputs a planning behavior.
\textbf{Red--green watermarking (RG)} \cite{huang2025agent} is a bias-based baseline. At each step, we deterministically partition $\mathcal{B}_t$ into a green set of size $\gamma|\mathcal{B}_t|$ using a keyed PRG ($\gamma=0.5$). We add a logit bias $\delta$ to green behaviors before sampling ($\delta=2.0$). We detect the watermark by testing whether selected behaviors fall into the green set more often than the $\gamma$ baseline. \textbf{AgentMark-F (Ours)} samples $\hat b_t$ via distribution-preserving watermark sampling on $P_t$. Both AgentMark-F and RG use the same keyed-PRG design to derive reproducible per-step randomness.

\subsection{Utility and Capacity}
\label{sec:exp-utility-capacity}

\paragraph{Embodied and tool-use benchmarks.}
From the results in Table~\ref{tab:main_results}, we can get the following conclusion. Firstly, RG degrades utility on long-horizon embodied tasks (ALFWorld-ID SR $89.5\%\!\rightarrow\!78.8\%$; steps $19.7\!\rightarrow\!26.1$). Secondly, AgentMark-F preserves utility across embodied and tool-use settings (close to baseline on ALFWorld ID/OOD and ToolBench) while achieving non-trivial capacity (ALFWorld $\sim$1.2--1.3 bps; ToolBench $\sim$0.49 bps, 4.93 bpt). Finally, the per-step overhead is small in both latency and token cost: $\Delta$s/step is near zero and $\Delta$Tok/step stays within $\pm 0.5\%$ on ALFWorld, while ToolBench shows no systematic increase and is often lower due to earlier termination (Avg.\ $\Delta$Tok/step $=-6.25\%$). Additional setup details are provided in Appendix~\ref{appendix:utility-capacity-ALFWorld-ToolBench}.
Appendix~\ref{sec:no_induction} first checks that the elicitation interface itself does not materially change task success. We further validate these findings on Gemini 2.0 Flash in Appendix~\ref{sec:cross_model}, confirming model-agnostic utility preservation. Complementary step-level capacity--entropy analysis is provided in Appendix~\ref{sec:capacity_entropy}.

\begin{figure}[t]
  \centering
  \includegraphics[width=\columnwidth]{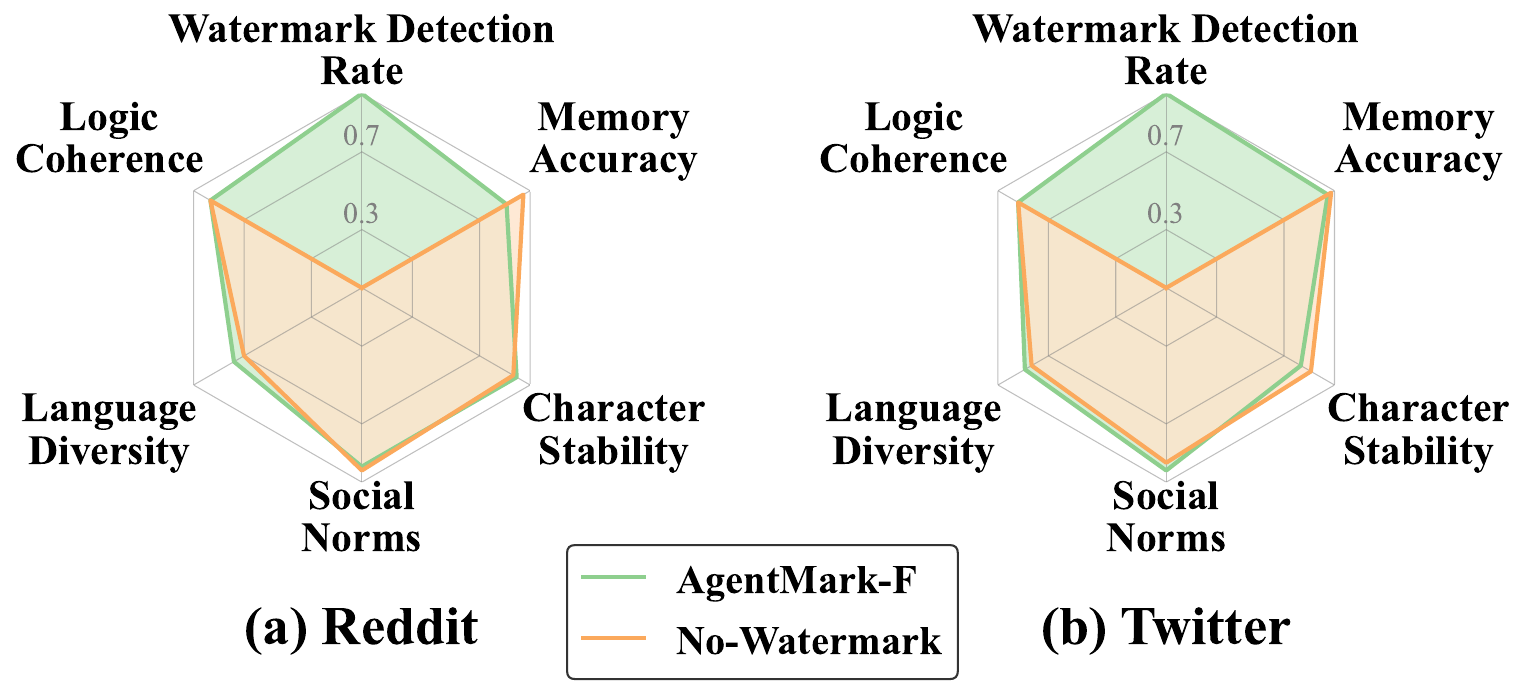}
  \caption{OASIS social-quality utility and detectability.}
  \label{fig:oasis_radar}
\end{figure}
\vspace{-0.8em}
\paragraph{Social simulation.}
We evaluate deployment feasibility in OASIS on Reddit- and Twitter-like platforms, running 100 trajectories per platform with ours and a no-watermark control split 50/50.
We score social-quality utility with an LLM judge on five dimensions (coherence, memory accuracy, character stability, social norms/common sense, and language diversity) and report Watermark Detection Rate as decoding success from behaviors.
We simulate both platforms under matched seeds and scenario scripts so that differences are attributable to watermarking rather than environmental variation.
This setting tests whether behavioral provenance remains recoverable while preserving persona consistency and socially appropriate interactions in open-ended multi-agent feeds.
Figure~\ref{fig:oasis_radar} shows that ours preserves social-quality utility while maintaining high watermark verifiability; details are in Appendix~\ref{appendix:utility-capacity-oasis}.

\subsection{Robustness Experiment}

\begin{figure}[t]
  \centering
  \includegraphics[width= \columnwidth]{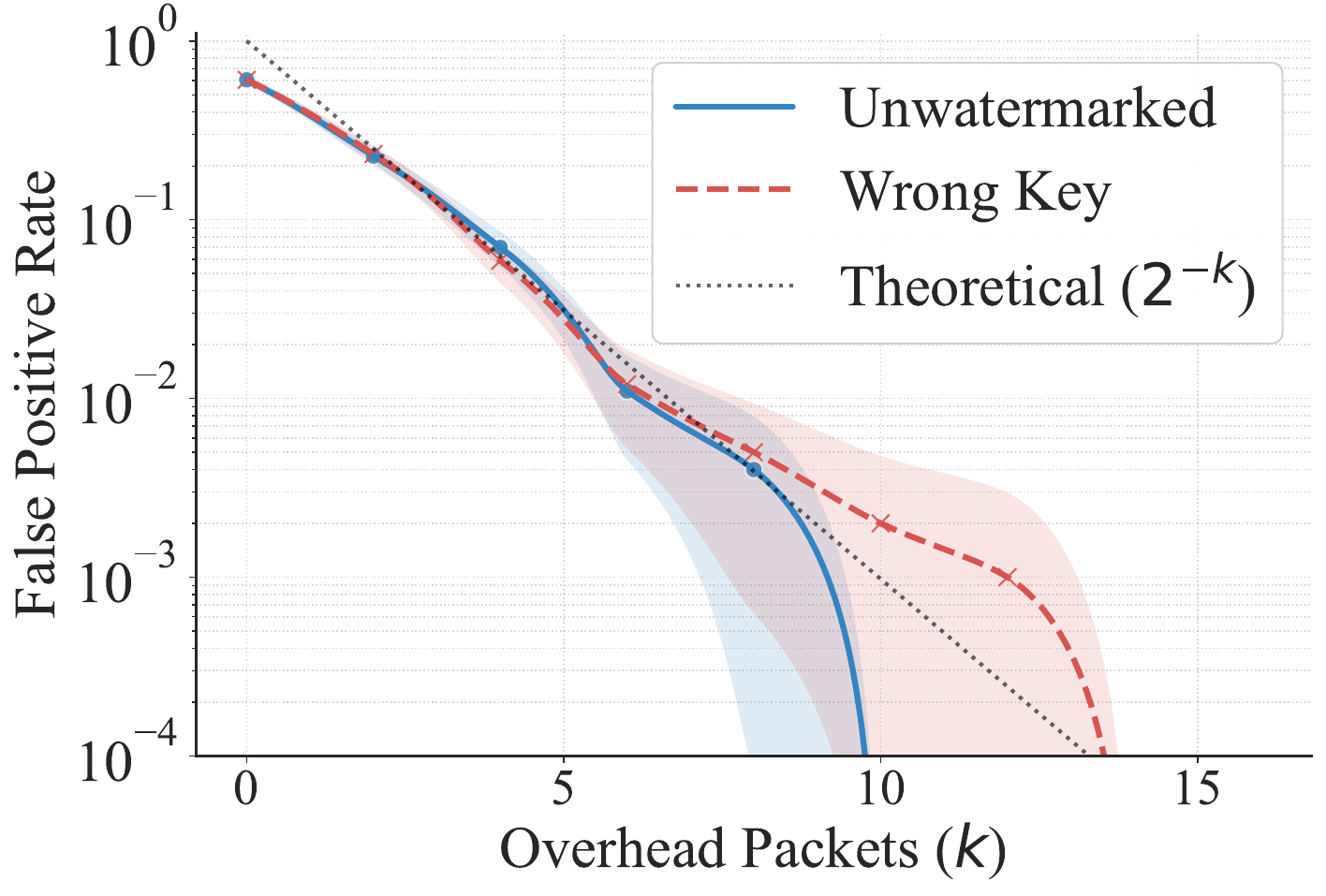}
  \caption{Both unwatermarked and wrong-key FPRs decay as $2^{-k}$ against the overhead $k$.}
  \label{fig:fpr}
\end{figure}

\paragraph{False Positives and Key Forgery}
We run 1000 Monte Carlo trials per overhead packet $k\in[0,16]$ with payload length $N=128$, using 281 ToolBench trajectories (T1--T3) and 159 unique action indices to construct the coefficient matrix (Appendix~\ref{appendix:fpr}).
Figure~\ref{fig:fpr} reports the verifier's false-positive rates (FPR) under unwatermarked logs and wrong-key decoding, where verification accepts only if the induced GF(2) system is consistent.
FPR drops below $1\%$ for $k\ge 8$, and we observe no false positives for $k\ge 14$.
Both curves decay exponentially with $k$, closely tracking the $2^{-k}$ trend predicted by Eqs.~\eqref{eq:rlnc-rank-bound} and~\eqref{eq:rlnc-success-lb}.

\begin{figure}[t]
  \centering
  \includegraphics[width=\columnwidth]{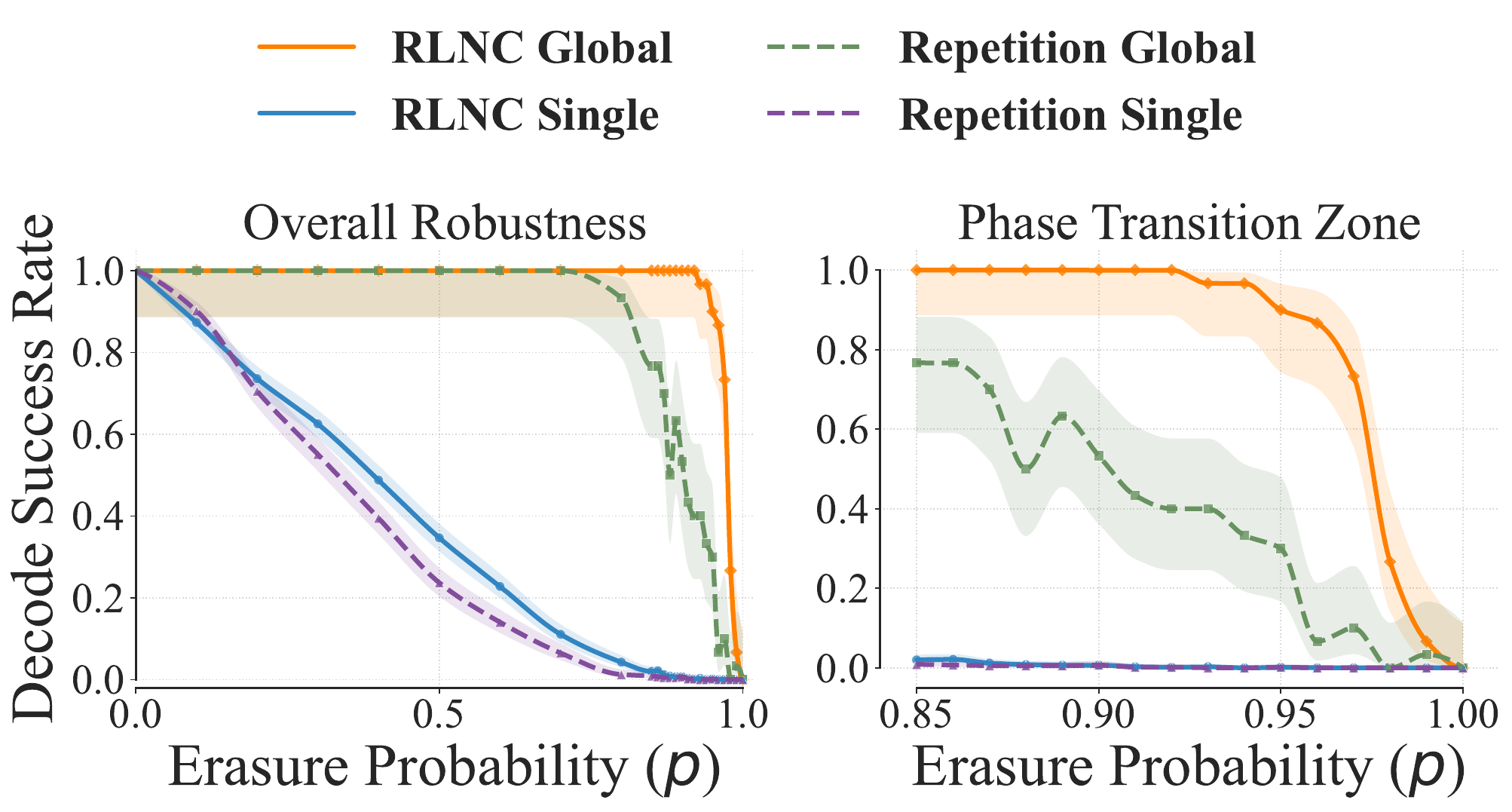}
  \caption{Robustness to Step Erasure and Truncation.}
  \label{fig:rlnc_comparison}
\end{figure}

\paragraph{Robustness to Step Erasure and Truncation.}
We evaluate robustness to step erasure by independently dropping each logged decision step with probability $p$ and decoding from the remaining records.
We compare RLNC-based coding with a repetition baseline, under single-episode decoding (per trajectory) and global decoding (aggregation across trajectories); details are in Appendix~\ref{appendix:robustness}.
Figure~\ref{fig:rlnc_comparison} shows that RLNC is substantially more erasure-resilient than repetition, and global aggregation further improves robustness by accumulating enough independent equations for full-rank recovery, with a clear phase transition only when the effective packet budget becomes insufficient.

\paragraph{Robustness to Semantic-Preserving Observation Rewriting.}
Table~\ref{tab:rewrite_sensitivity} reports a high-pressure stress test that semantically rewrites step observations while keeping the task state unchanged, emulating deployments where raw logs are missing and only semantically equivalent reconstructions are available for verification; from these results, we can get the following conclusion.
First, rewriting can noticeably shift the elicited behavior distribution (Avg.\ KL $=3.227\pm 1.802$), so perfect step-level synchronization is not expected.
Second, the moderate mean divergence with substantial variance indicates that many steps remain stable and yield similar $P_t$, preserving aligned decisions and recoverable watermark signal (Behavior Match Rate $=49.45\pm 16.90\%$, Bit Recovery Rate $=16.84\pm 19.56\%$).
Finally, since ID payloads are always short while agent executions are long-horizon, aggregation over surviving aligned steps can still support practical verification.
Further details are provided in Appendix~\ref{appendix:semantic}.
We further extend the robustness analysis to cross-run distribution stability in Appendix~\ref{sec:jsd_stability} and discuss adaptive threat models in Appendix~\ref{sec:adaptive_adversary}.

\begin{table}[t]
\centering
\setlength{\tabcolsep}{7pt}
\renewcommand{\arraystretch}{1.1}
\resizebox{\columnwidth}{!}{%
\begin{tabular}{lccc}
\toprule
Condition &
\makecell{Behavior Match\\Rate (\%) $\uparrow$} &
\makecell{Avg. KL} &
\makecell{Bit Recovery\\Rate (\%) $\uparrow$} \\
\midrule
No rewriting & 100.00 & 0.000 & 100.00 \\
Semantic rewriting     & $49.45 \pm 16.90$ & $3.227 \pm 1.802$ & $16.84 \pm 19.56$ \\
\bottomrule
\end{tabular}%
}
\caption{Sensitivity to semantic-preserving observation rewriting on ALFWorld-OOD (134 tasks; 2326 steps). Behavior Match Rate is the fraction of steps whose planning behavior matches between original and rewritten observations. Avg.\ KL is the mean KL divergence between the elicited behavior distributions. Bit Recovery Rate is the fraction of watermarked steps whose extracted bitstring matches between the two observations.}
\label{tab:rewrite_sensitivity}
\end{table}

\subsection{Content-Watermark Compatibility}
As summarized in Table~\ref{tab:compat}, AgentMark watermarks planning behaviors and is complementary to action-layer content watermarking.
We evaluate composability with distribution-preserving SynthID-Text by enabling both watermarks and show that task utility is preserved, behavioral decoding for Ours remains correct (100\%), and SynthID-Text detection under Both stays high (96.6\% on ToolBench; details in Appendix~\ref{appendix:compat}).
Overall, the two watermarks provide complementary provenance under different failures: AgentMark-F targets trajectory-level log erasure and truncation, whereas SynthID-Text targets robustness of final content under rewriting.

\begin{table}[t]
\centering
\resizebox{\columnwidth}{!}{%
\begin{tabular}{lcccc}
\toprule
Config & \makecell{Multi\\bit} & \makecell{Behav.\\cap.} & \makecell{Content\\det.} & \makecell{Threat\\model} \\
\midrule
SynthID-Text      & \ding{55} & --   & \ding{51} & \makecell{rewrite} \\ \midrule
AgentMark-F & \ding{51} & high & --        & \makecell{erasure} \\ \midrule
Both         & \ding{51} & high & \makecell{\ding{51}} & \makecell{rewrite\\\& erasure} \\
\bottomrule
\end{tabular}%
}
\caption{Composability of AgentMark-F and SynthID.}
\label{tab:compat}
\end{table}

\section{Conclusion}
We propose AgentMark, which embeds multi-bit identifiers into behaviors via distribution-preserving sampling under black-box APIs, supporting recovery under step erasure and truncation while preserving utility.
Across embodied, tool-use, and social environments, we show non-trivial capacity, reliable decoding from partial logs, and compatibility with existing action-layer content watermarking.
We hope AgentMark makes provenance a deployable primitive for scalable auditing and accountability, thereby laying the groundwork for AI governance and safeguarding societal security in the era of autonomous agents.

\section*{Limitations}
AgentMark requires the agent to output an explicit planning-time behavior distribution $P_t$ over a predefined behavior list and to log planning-time decisions (and minimal step context) for verification; deployments that cannot retain sufficient logs may limit provenance recovery. Since verification relies on step observations to reproduce sampling randomness, semantic variation in observations (e.g., paraphrasing) may reduce synchronization and thus degrade verification quality. Because AgentMark preserves the original behavior distribution, the per-step embedding capacity is naturally context-dependent and can be small when $P_t$ is highly peaked, which may reduce per-episode capacity without aggregation. Finally, our approach exploits the fact that an LLM induces an implicit conditional policy over a finite behavior set, and makes this policy auditable by eliciting an explicit probability list, which enables provenance under black-box APIs; for open-source LLMs, an important future direction is to extract such planning-time policies directly from logits or latent representations, thereby enabling provenance without explicit elicitation.

\section*{Acknowledgements}
The authors are immensely grateful to Guorui Liao for his vital mentorship and insightful discussions on provable security, which significantly shaped this research. This work was supported in part by the National Key Research and Development Program of China under Grant 2023YFC3305402 and in part by the National Natural Science Foundation of China under Grant 62172053 and Grant 62302059.

\bibliography{custom}

\appendix

\section{Deterministic Differential Recombination}
\label{appendix:recombination}
\begin{figure}[t]
  \centering
  \includegraphics[width=\columnwidth]{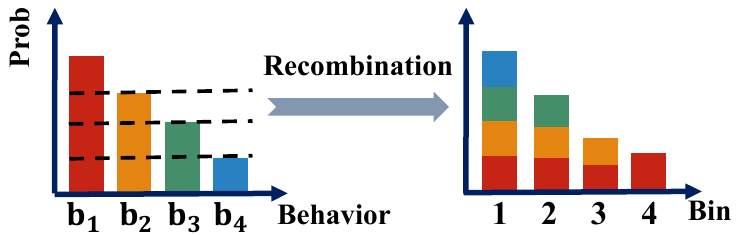}
  \caption{Differential-based recombination slices the behavior distribution into equal-probability layers and recombines them into a mixture of uniform bins.}
  \label{fig:Diff}
\end{figure}

This appendix provides an explicit and deterministic procedure for the differential-based probability recombination used by AgentMark-F (Figure~\ref{fig:Diff}).

The goal is to transform an arbitrary discrete distribution $P_t$ over a finite set $\mathcal{B}_t$ into a \emph{mixture of uniform bins} that preserves the original marginal distribution while enabling uniform encoding within a selected bin.
Our construction is an instantiation of the probability recombination module in FDPSS. For the general framework and security discussions, see \cite{liao2025framework}.

\paragraph{Setup and notation.}
Let $\mathcal{B}_t=\{b_{t,1},\ldots,b_{t,n}\}$ be the candidate behavior set with probabilities $p_i = P_t(b_{t,i})$.
We assume $\sum_{i=1}^n p_i = 1$ and $p_i \ge 0$.
To simplify notation, we define $p_{n+1}=0$.

\paragraph{Canonical ordering under probability ties.}
Differential recombination relies on a sorted ordering of behaviors by probability.
In practice, multiple behaviors may have identical probabilities (or probabilities that are equal up to numerical precision).
To ensure encoder--decoder synchronization, we define a canonical ordering as follows. We first quantize probabilities to a fixed precision, e.g., $p_i \mapsto \mathrm{round}(p_i;\pi)$ for a chosen precision $\pi$. We then apply a stable sort by the quantized probabilities in non-increasing order, so that behaviors with equal (quantized) probabilities preserve a consistent relative order across runs.
This produces an ordered sequence $(b_{t,1},\ldots,b_{t,n})$ such that $p_1 \ge \cdots \ge p_n$ with deterministic behavior under ties.

\paragraph{Differential slicing and bin construction.}
The differential-based scheme can be understood as horizontal slicing of the sorted probability histogram.
Define slice heights
\begin{equation}
d_k = p_k - p_{k+1}, \qquad k \in \{1,\ldots,n\}.
\end{equation}
Each slice $k$ has height $d_k$ and spans the top-$k$ behaviors.
Equivalently, each behavior $b_{t,i}$ is decomposed into blocks $\{d_k\}_{k\ge i}$ since
\begin{equation}
p_i = \sum_{k=i}^{n} d_k.
\end{equation}
We group the $k$ equal-height blocks at level $d_k$ into a \emph{uniform bin} $T_k=\{b_{t,1},\ldots,b_{t,k}\}$.
The mixture weight (total probability mass) of bin $k$ is
\begin{equation}
q_k = k \cdot d_k.
\end{equation}
Bins with $d_k=0$ have zero weight and can be omitted.

\paragraph{Marginal distribution preservation.}
If we first sample a bin index $K \sim \mathrm{Cat}(q_1,\ldots,q_n)$ and then sample uniformly within the selected bin $T_K$, the marginal probability of selecting behavior $b_{t,i}$ is:
\begin{equation}
\begin{aligned}
\Pr[\hat b_t=b_{t,i}]
&= \sum_{k=i}^{n} q_k \cdot \frac{1}{k} \\
&= \sum_{k=i}^{n} (p_k - p_{k+1}) \\
&= p_i ,
\end{aligned}
\end{equation}
which matches the original distribution.

\paragraph{Algorithm.}
Algorithm~\ref{alg:diff-recombine} summarizes the deterministic recombination procedure.
It outputs the ordered behaviors and the corresponding list of non-zero bins with mixture weights, which are then used by the bin-sampling and uniform-encoding modules.

\begin{algorithm}[t]
\caption{Deterministic differential-based recombination}
\label{alg:diff-recombine}
\begin{algorithmic}[1]
\Require Candidate set $\mathcal{B}_t=\{b_1,\ldots,b_n\}$, distribution $P_t$, quantization precision $\pi$
\Ensure Ordered list $(b_{(1)},\ldots,b_{(n)})$, bins $\{T_k\}$ and weights $\{q_k\}$
\State $p_i \gets P_t(b_i)$ for all $i$; $p_{n+1}\gets 0$
\State $\tilde p_i \gets \mathrm{round}(p_i;\pi)$ for all $i$ \Comment{quantize probabilities}
\State $(b_{(1)},\ldots,b_{(n)}) \gets \mathrm{StableSort}(\{b_i\};~\tilde p_i~\text{descending})$
\State $p_k \gets P_t(b_{(k)})$ for $k=1,\ldots,n$; $p_{n+1}\gets 0$
\For{$k=1$ to $n$}
  \State $d_k \gets p_k - p_{k+1}$
  \If{$d_k > 0$}
    \State $T_k \gets \{b_{(1)},\ldots,b_{(k)}\}$ \Comment{top-$k$ behaviors}
    \State $q_k \gets k \cdot d_k$
  \EndIf
\EndFor
\State \Return $(b_{(1)},\ldots,b_{(n)}), \{(T_k,q_k): d_k>0\}$
\end{algorithmic}
\end{algorithm}

\section{Cyclic-Shift Uniform Encoding and Decoding}
\label{appendix:cyclicshift}

This appendix details the cyclic-shift uniform encoding/decoding module used by AgentMark-F within a selected uniform bin.
The role of this module is to embed a (variable-length) prefix of the payload bitstream into a uniform choice among $n$ candidates, while remaining invertible given the same pseudorandomness.
Our presentation follows the standard uniform-steganography component used in FDPSS-style constructions; see \cite{liao2025framework} for further discussion and security proofs.

\paragraph{Setting.}
Fix a time step $t$ and a selected bin $T$ containing $n = |T|$ behaviors, which is treated as a uniform distribution over indices $\{0,1,\ldots,n-1\}$.
Let $M$ denote the payload bitstream and $\ell$ the current pointer, so the available suffix is $M[\ell:]$.
Both encoder and decoder have synchronized access to a pseudorandom generator $\mathsf{PRG}$ (seeded as described in the main text), from which they draw a shared random value that induces a cyclic shift.

\paragraph{Key idea.}
Let $k \triangleq \lfloor \log_2 n \rfloor$ and $t \triangleq n - 2^k$, so $n = 2^k + t$ with $0 \le t < 2^k$.
CyclicShift embeds either $k$ bits or $k{+}1$ bits depending on the payload prefix, yielding near-optimal utilization of the uniform bin.
Intuitively, $2^k-t$ indices correspond to $k$-bit codewords and the remaining $2t$ indices correspond to $(k{+}1)$-bit codewords, forming a prefix-free set of size $n$.

\paragraph{Shared cyclic shift.}
Both sides draw $\mathrm{ptr}\in[0,1)$ from $\mathsf{PRG}$ and compute a shift
\begin{equation}
R \triangleq \lfloor \mathrm{ptr}\cdot n \rfloor \in \{0,1,\ldots,n-1\}.
\end{equation}
Because $\mathsf{PRG}$ is synchronized, the decoder reconstructs the same $R$.

\paragraph{Algorithms.}
Algorithm~\ref{alg:cyclic-enc} and Algorithm~\ref{alg:cyclic-dec} give the one-step cyclic-shift encoding and decoding procedures.
The encoder outputs an index $j$ within a size-$n$ bin and the embedded substring $s$ (either $k$ or $k{+}1$ bits).
The decoder takes the observed index $j$ and recovers the same substring $s$.

\paragraph{Correctness.}
For $n>1$, both algorithms use the same pseudorandom shift $R$ and apply inverse mappings between payload prefixes and bin indices, hence decoding inverts encoding at each step and recovers the embedded substring $s$ exactly.
For $n=1$, there is no choice within the bin and the embedded bitstring is empty.

\paragraph{Uniformity.}
The cyclic shift randomizes the mapping between payload prefixes and indices.
When combined with a pseudorandom payload bitstream (e.g., after encryption/coding as assumed in standard steganography formulations), the induced selection within a uniform bin is indistinguishable from uniform sampling.
We refer readers to \cite{liao2025framework} for formal security statements in the FDPSS setting.

\paragraph{Capacity (near-optimality).}
For a bin of size $n$, CyclicShift embeds a variable number of bits $c\in\{k,k{+}1\}$ where $k=\lfloor \log_2 n\rfloor$ and $n=2^k+m$ with $0\le m<2^k$ (and $c=0$ when $n=1$).
The prefix-free codebook induced by CyclicShift assigns $2^k-m$ indices to $k$-bit codewords and $2m$ indices to $(k{+}1)$-bit codewords, so under a uniform choice among $n$ indices the expected embedded length is
\begin{equation}
\begin{aligned}
\mathbb{E}[c\mid n]
&= \frac{(2^k-m)\cdot k + (2m)\cdot (k+1)}{n} \\
&= k+\frac{2m}{n}.
\end{aligned}
\end{equation}
This yields the information-theoretic upper bound $\mathbb{E}[c\mid n]\le \log_2 n$.
Moreover, writing
\begin{equation}
\mathbb{E}[c\mid n]
= \log_2 n - \log_2\!\left(1+\frac{m}{2^k}\right) + \frac{2m}{n},
\end{equation}
and letting $x\triangleq m/2^k\in[0,1)$, we have $\log_2(1+x)-x \le 0.0861$ for all $x\in[0,1)$, which implies the constant-gap lower bound
\begin{equation}
\mathbb{E}[c\mid n]\ge \log_2 n - 0.0861.
\end{equation}
Therefore, CyclicShift achieves near-optimal expected capacity for uniform bins.

\begin{algorithm}[t]
\caption{CyclicShift (one step): \textsc{Encode}}
\label{alg:cyclic-enc}
\begin{algorithmic}[1]
\Require Payload suffix $M[\ell:]$, bin size $n$, synchronized $\mathsf{PRG}$
\Ensure Index $j\in\{0,\ldots,n-1\}$, embedded bits $s$
\If{$n=1$}
  \State \Return $0,\ \emptyset$
\EndIf
\State $\mathrm{ptr} \gets \mathsf{PRG}()$; $R \gets \lfloor \mathrm{ptr}\cdot n \rfloor$
\State $k \gets \lfloor \log_2 n \rfloor$; $t \gets n-2^k$
\If{$|M[\ell:]|  < k$}
\State \Return $R,\ \emptyset$
\EndIf
\State $x \gets \mathrm{bits2int}(M[\ell:\ell+k])$
\State $r \gets \begin{cases} M[\ell+k] & \text{if } |M[\ell:]|\ge k+1 \\ 0 & \text{otherwise} \end{cases}$
\If{$x < 2^k - t$}
  \State $j \gets (x+R)\bmod n$; $s \gets M[\ell:\ell+k]$
\Else
  \State $j \gets \big(2(x-(2^k-t)) + (2^k-t) + R + r\big)\bmod n$
  \State $s \gets M[\ell:\ell+k+1]$
\EndIf
\State \Return $j,\ s$
\end{algorithmic}
\end{algorithm}

\begin{algorithm}[t]
\caption{CyclicShift (one step): \textsc{Decode}}
\label{alg:cyclic-dec}
\begin{algorithmic}[1]
\Require Observed index $j\in\{0,\ldots,n-1\}$, bin size $n$, synchronized $\mathsf{PRG}$
\Ensure Extracted bits $s$ (possibly empty)
\If{$n=1$}
  \State \Return $\emptyset$
\EndIf
\State $\mathrm{ptr} \gets \mathsf{PRG}()$; $R \gets \lfloor \mathrm{ptr}\cdot n \rfloor$
\State $k \gets \lfloor \log_2 n \rfloor$; $t \gets n-2^k$
\State $\mathrm{idx} \gets (j-R)\bmod n$
\If{$\mathrm{idx} < 2^k - t$}
  \State $s \gets \mathrm{int2bits}(\mathrm{idx},k)$
\Else
  \State $u \gets \mathrm{idx}-(2^k-t)$
  \State $x \gets \lfloor u/2 \rfloor + (2^k-t)$
  \State $r \gets u \bmod 2$
  \State $s \gets \mathrm{int2bits}(x,k)\ \Vert\ r$
\EndIf
\State \Return $s$
\end{algorithmic}
\end{algorithm}

\section{End-to-End Example}
\label{appendix:example}

\paragraph{One-step example (ticket purchase).}
We illustrate a single watermark-embedding step for an agent assisting with ticket purchase.
Suppose the user goal is to buy a ticket, and the step-$t$ observation is
\begin{equation}
\begin{aligned}
o_t = \texttt{User: book a ticket.}
\end{aligned}
\end{equation}
with step index $t=7$ and shared secret $K_{\mathrm{sh}}$.
For concreteness, we instantiate the step context as a short interaction trace,
\begin{equation}
\begin{aligned}
h_t = [&\texttt{(}t{-}2,\ \textsf{Search}\texttt{)},\ \texttt{(}t{-}1,\ \textsf{Book}\texttt{)},\\
        &\texttt{User preference: window seat.},\\
        &\texttt{Budget: \$300.}]
\end{aligned}
\end{equation}
which is available to both the encoder and verifier from the same logs.
We therefore write the shared step context as $Context_t \triangleq (t,o_t,h_t)$.

The agent elicits the following behavior distribution $P_t$ over
$\mathcal{B}_t$=\{Search,Book,Pay,Check-in,Modify\}:
\begin{equation}
P_t = (0.40,\,0.25,\,0.15,\,0.12,\,0.08),
\end{equation}
which sums to $1$.

\paragraph{Differential recombination.}
After sorting in non-increasing order (already in the above order), we have
\(p_1=0.40,p_2=0.25,p_3=0.15,p_4=0.12,p_5=0.08\), and \(p_6=0\).
The slice heights are
\begin{equation}
(d_1,d_2,d_3,d_4,d_5)=(0.15,\,0.10,\,0.03,\,0.04,\,0.08).
\end{equation}
Differential recombination forms uniform bins $T_k$ (top-$k$ behaviors) with mixture weights $q_k=k\cdot d_k$:
\begin{equation}
(q_1,q_2,q_3,q_4,q_5)=(0.15,\,0.20,\,0.09,\,0.16,\,0.40),
\end{equation}
and
\begin{equation}
\begin{aligned}
T_1 &= \{\textsf{Search}\},\\
T_2 &= \{\textsf{Search},\textsf{Book}\},\\
T_3 &= \{\textsf{Search},\textsf{Book},\textsf{Pay}\},\\
T_4 &= \{\textsf{Search},\textsf{Book},\textsf{Pay},\textsf{Check\mbox{-}in}\},\\
T_5 &= \{\textsf{Search},\textsf{Book},\textsf{Pay},\textsf{Check\mbox{-}in},\textsf{Modify}\}.
\end{aligned}
\end{equation}

\paragraph{Keyed pseudorandomness and bin selection.}
We derive the per-step key $K_t \leftarrow H(K_{\mathrm{sh}}\|Context_t)$ and initialize $\mathsf{PRG}\gets\mathsf{PRG}(K_t)$.
Assume the first pseudorandom draw for bin selection is $u_1=0.62$. With cumulative weights $(0.15,0.35,0.44,0.60,1.00)$, this selects $K=5$, hence the selected bin is $T=T_5$ with size $n=|T|=5$.

\paragraph{Payload embedding via CyclicShift within the bin.}
Suppose the next payload bits to embed are $m_t=\texttt{01}$.
To pseudorandomize the embedded bits, we generate a pad from the same per-step key and bin-dependent length:
\begin{equation}
Z_t \leftarrow \mathsf{PRG}(K_t), \qquad \tilde m_t \triangleq m_t \oplus Z_t[1{:}2].
\end{equation}
For example, if $Z_t[1{:}2]=\texttt{11}$ then $\tilde m_t=\texttt{10}$.
CyclicShift then uses a second pseudorandom draw $u_2=0.27$ to set the cyclic shift
\(R=\lfloor u_2\cdot n\rfloor=\lfloor 0.27\cdot 5\rfloor=1\).

With $n=5$, we have $k=\lfloor \log_2 5\rfloor=2$ and $n-2^k=1$.
Interpreting the first $k$ bits of $\tilde m_t$ as $x=\mathrm{bits2int}(\texttt{10})=2$, since $x<2^k-(n-2^k)=3$, CyclicShift selects index
\begin{equation}
j=(x+R)\bmod 5=(2+1)\bmod 5=3,
\end{equation}
and embeds $s_t=\texttt{10}$ (two bits).
Therefore the watermarked behavior is the $j$-th element in $T_5$, i.e.,
\begin{equation}
\hat b_t = T_5[3] = \textsf{Check\mbox{-}in}.
\end{equation}
Although this particular realization selects \textsf{Check-in}, the marginal distribution over $\hat b_t$ remains $P_t$ by construction.

\paragraph{Decoding (same step).}
Given $Context_t$ and the observed behavior $\hat b_t=\textsf{Check-in}$, the decoder derives the same $K_t$, reproduces the same bin $T_5$ and shift $R=1$, and recovers the same index $j=3$.
CyclicShift decoding returns $s_t=\texttt{10}$, and the original payload bits are recovered by unmasking:
\begin{equation}
m_t = \tilde m_t \oplus Z_t[1{:}2] = \texttt{10}\oplus \texttt{11}=\texttt{01}.
\end{equation}

\paragraph{Takeaway.}
This one-step example shows how AgentMark-F embeds provenance bits into behavior selection while preserving the original behavior distribution.
By enabling reliable attribution from logged behaviors, AgentMark provides a provenance mechanism that does not impair task utility, making it practical for agent providers to deploy in real systems. This supports protecting intellectual property against unauthorized cloning or misuse and offers a concrete tool for auditing and accountability in regulated or platform-mediated settings.

\section{Security Proof}
\label{appendix:Security_Proof}
\paragraph{Distribution-preserving watermark embedding.}
AgentMark-F combines differential recombination and cyclic-shift uniform encoding to embed provenance bits while preserving the per-step behavior distribution.
First, differential recombination rewrites an arbitrary distribution $P_t$ as a mixture of uniform bins without changing its marginal.
Next, within the selected uniform bin, CyclicShift embeds bits by choosing an index in a way that is (computationally) indistinguishable from uniform sampling.
Finally, we pseudorandomize the embedded payload bits at each step using the per-step key, so that the bits presented to CyclicShift are pseudorandom even if the original payload has structure.

\paragraph{Pseudorandomizing the payload bits.}
\label{appendix:Pseudorandomizing}
Let $K_{\mathrm{sh}}$ be the shared secret and let $Context_t$ denote the shared step context at step $t$ (e.g., the step index, observation, and logged history available to both sides).
We derive a per-step key
\begin{equation}
K_t \leftarrow H(K_{\mathrm{sh}} \,\|\, Context_t),
\end{equation}
and use a keyed pseudorandom generator to produce a pad bitstream
\begin{equation}
Z_t \leftarrow \mathsf{PRG}(K_t) \in \{0,1\}^{\infty}.
\end{equation}
At step $t$, CyclicShift embeds a variable number of bits $c_t \in \{k, k{+}1\}$ where $k=\lfloor \log_2 |T|\rfloor$ and $|T|$ is the selected bin size.
Let $m_t \in \{0,1\}^{c_t}$ be the next $c_t$ payload bits to embed.
We mask them with the per-step pad:
\begin{equation}
\tilde m_t \triangleq m_t \oplus Z_t[1{:}c_t],
\end{equation}
where $\oplus$ denotes a generic keyed pseudorandom mixing operation (e.g., bitwise XOR as a canonical instantiation), so long as $\tilde m_t$ is computationally indistinguishable from uniform given $K_t$ and the corresponding pad segment.

The encoder feeds $\tilde m_t$ to CyclicShift to choose the bin index, while the decoder recovers $\tilde m_t$ and unmask it using the same pad bits.
Under standard PRG assumptions, $\tilde m_t$ is computationally indistinguishable from uniform even if $m_t$ is not.

\paragraph{Marginal distribution preservation.}
Let $\mathcal{B}_t=\{b_{t,1},\ldots,b_{t,n}\}$ be ordered so that $p_i=P_t(b_{t,i})$ and $p_1\ge\cdots\ge p_n$, with $p_{n+1}=0$.
Differential recombination defines $d_k=p_k-p_{k+1}$ and mixture weights $q_k=k\cdot d_k$, where bin $T_k=\{b_{t,1},\ldots,b_{t,k}\}$ is uniform over its $k$ elements.
Sampling $K\sim \mathrm{Cat}(q_1,\ldots,q_n)$ and then selecting uniformly from $T_K$ preserves the marginal:
\begin{equation}
\begin{aligned}
\Pr[\hat b_t=b_{t,i}]
&= \sum_{k=i}^{n} q_k \cdot \frac{1}{k} \\
&= \sum_{k=i}^{n} (p_k - p_{k+1}) \\
&= p_i .
\end{aligned}
\end{equation}

\paragraph{Putting the pieces together.}
By the identity above, the recombination-based sampler is equivalent to drawing from $P_t$ at each step.
Replacing the ideal randomness used in bin selection and cyclic-shift encoding by $\mathsf{PRG}(K_t)$ yields a computationally indistinguishable sampling process under standard PRG security.
Masking the payload bits by $\tilde m_t = m_t \oplus Z_t[1{:}c_t]$ ensures that the input bits consumed by CyclicShift are pseudorandom, aligning with the uniform-steganography requirement in FDPSS-style analyses \cite{liao2025framework}.
Therefore, AgentMark-F embeds provenance bits while keeping the induced behavior distribution computationally indistinguishable from sampling $b_t\sim P_t$.

\section{RLNC Details and Advantages}
\label{appendix:rlnc}

This appendix formalizes why RLNC is a natural fit for variable-capacity behavioral watermarking under erasure/truncation and provides decoding details.
We refer to \cite{ho2006random} for classical RLNC analysis.

\paragraph{Variable-capacity embedding as a rateless linear system.}
At step $t$, the sampler outputs $s_t\in\{0,1\}^{c_t}$ where the capacity $c_t$ varies with the selected bin size.
We treat each embedded bit as one linear equation over $\mathbb{F}_2$:
\begin{equation}
y_{t,j} = \langle a_{t,j}, m\rangle,\qquad a_{t,j}\in\mathbb{F}_2^{L}, j=1,\ldots,c_t,
\end{equation}
so each step contributes exactly $c_t$ equations.
Under erasure/truncation, the verifier observes $\mathcal{I}\subseteq\{1,\ldots,T\}$ and receives $R=\sum_{t\in\mathcal{I}}c_t$ equations, yielding a linear system
\begin{equation}
A_{\mathcal{I}}\, m = y_{\mathcal{I}},\qquad A_{\mathcal{I}}\in\mathbb{F}_2^{R\times L},\ y_{\mathcal{I}}\in\mathbb{F}_2^{R}.
\end{equation}
This formulation is \emph{rateless} in the sense that decoding depends only on how many equations are ultimately collected, not on a fixed per-step rate.

\paragraph{Erasure robustness via rank.}
Unique recovery holds iff $\mathrm{rank}(A_{\mathcal{I}})=L$.
If the coefficient rows behave as (pseudo)random in $\mathbb{F}_2^{L}$, then for any realization with $R=L+\Delta$ equations,
\begin{equation}
\Pr[\mathrm{rank}(A_{\mathcal{I}})=L] \ge 1-2^{-\Delta},
\end{equation}
which yields an exponentially decaying failure probability as additional equations accrue beyond $L$.
Crucially, this bound is agnostic to \emph{which} steps are erased: only the total received capacity $R=\sum_{t\in\mathcal{I}}c_t$ matters.

\paragraph{Keyed pseudorandom coefficient generation.}
To make coefficients reproducible for the verifier while remaining pseudorandom, we derive a per-step key
\begin{equation}
K_t \leftarrow H(K_{\mathrm{sh}} \,\|\, Context_t),
\end{equation}
and generate coefficients deterministically as
\begin{equation}
a_{t,j} \leftarrow \mathsf{PRG}(K_t,j)\in\mathbb{F}_2^{L}.
\end{equation}
Under standard PRG assumptions, these coefficients are computationally indistinguishable from uniform while being reproducible from $Context_t$ on the verifier side.

\paragraph{Decoding algorithm (Gaussian elimination over $\mathbb{F}_2$).}
Given $(A_{\mathcal{I}},y_{\mathcal{I}})$, the verifier solves for $m$ via Gaussian elimination in $\mathbb{F}_2$.
The time complexity is $O(L^3)$ in the worst case for dense matrices, with substantially lower cost under sparsity or incremental updates.

\begin{algorithm}[t]
\caption{RLNC decode over $\mathbb{F}_2$ (sketch)}
\label{alg:rlnc-decode}
\begin{algorithmic}[1]
\Require Rows $(a_\ell,y_\ell)\in\mathbb{F}_2^{L}\times\mathbb{F}_2$ for $\ell=1,\ldots,R$
\Ensure Solution $m\in\mathbb{F}_2^{L}$ if full-rank; otherwise fail
\State Form matrix $A\in\mathbb{F}_2^{R\times L}$ and vector $y\in\mathbb{F}_2^{R}$
\State Perform Gaussian elimination on $[A\,|\,y]$ over $\mathbb{F}_2$
\If{$\mathrm{rank}(A)<L$}
  \State \Return fail
\Else
  \State \Return the unique solution $m$
\EndIf
\end{algorithmic}
\end{algorithm}

\paragraph{RLNC versus fixed-rate block codes.}
Block codes (e.g., Reed--Solomon\cite{reed1960polynomial}) typically assume fixed-size symbols and a predetermined codeword structure.
In our setting, the per-step capacity $c_t$ is variable and the observed trajectory may be an arbitrary subsequence $\mathcal{I}$, which makes fixed block boundaries and symbol alignment fragile.
RLNC avoids explicit block alignment by treating every embedded bit as a linear measurement and decoding from any subset of sufficiently many measurements, i.e., from any $\mathcal{I}$ that yields $R \ge L$ with high rank.

\section{Datasets and Benchmark Details}
\label{appendix:datasets}

\subsection{ALFWorld (ID/OOD)}
ALFWorld~\cite{shridhar2020alfworld} is a long-horizon embodied household benchmark built on interactive TextWorld environments aligned with the ALFRED setting~\cite{shridhar2020alfred}. Each episode requires multi-step planning and action execution to complete a goal (e.g., navigation and object interactions), making it a natural testbed for behavior-level watermarking where small distribution shifts can amplify over time. We report two evaluation regimes. The in-distribution (ID) split follows the standard test setting, and results are averaged over three independent runs on the same 100-task test set. The out-of-distribution (OOD) split evaluates on a 134-task cross-distribution test set with different task compositions. We report success rate (SR), average steps on successful episodes, and wall-clock time; watermark capacity is reported only for \textbf{Ours}.

\subsection{ToolBench Test Subsets}
ToolBench is a large-scale tool-use benchmark constructed from real-world APIs (e.g., RapidAPI), covering thousands of tools and tens of thousands of API endpoints, with diverse single-tool and multi-tool instructions that require structured tool calls and multi-step reasoning~\cite{liu2024agentbench}. We evaluate trajectories using ToolEval, an automatic evaluator that executes predicted tool calls (with bounded budgets) and scores whether the instruction is successfully completed. Following the standard ToolEval protocol, we report results on six test subsets that cover single-tool tool selection, single-tool instruction following, single-tool within-category selection, multi-tool within-category selection, multi-tool instruction following, and complex multi-tool instruction following.

\subsection{OASIS Platforms}
OASIS is a scalable social-media simulation environment that mimics platform dynamics of Twitter-like and Reddit-like ecosystems, where agents interact through diverse social actions (e.g., posting, commenting, reposting, following) under evolving feeds and recommendation mechanisms~\cite{yang2024oasis}. We simulate both platforms and compare watermarked and non-watermarked agents under the same seeds and scenario scripts, which enables a deployment-oriented assessment of behavioral consistency and social-quality utility. Beyond task-level logs for provenance, we evaluate social-quality dimensions including coherence, memory accuracy, persona consistency, social norms/common sense, and language diversity; we also report the average embedded bits per step as a reference statistic.

\section{Details of the Utility and Capacity Experiments}
\label{appendix:utility-capacity-experiments}

\subsection{Details in ALFWorld and ToolBench}
\label{appendix:utility-capacity-ALFWorld-ToolBench}

\paragraph{Evaluation.}
We evaluate task utility and watermark capacity/overhead under a unified agent execution pipeline across environments.
At each step, the agent outputs a planning behavior $b_t\in\mathcal{B}_t$ and then generates an execution action conditioned on $b_t$.
We compare three variants that differ only in how $b_t$ is selected: (i) \textbf{Baseline} follows the ReAct loop and directly outputs a planning behavior under the baseline policy $P_t^\star$; (ii) \textbf{RG} adds a logit bias to a pseudorandom green subset of $\mathcal{B}_t$ before sampling; and (iii) \textbf{AgentMark-F} elicits an explicit behavior distribution $P_t$ and samples $\hat b_t$ via distribution-preserving watermark sampling on $P_t$.

For each setting, we run three independent trials and report the mean and standard deviation.

\paragraph{ALFWorld and ToolBench details.}
We evaluate AgentMark-F and baselines on ALFWorld (ID/OOD splits) and ToolBench (six subsets covering single-tool and multi-tool instructions) under a unified agent interface. We use DeepSeek-Chat as the base model with temperature 1.0 on ALFWorld and 0.7 on ToolBench. On ALFWorld, we run 140 ID tasks and 134 OOD tasks for three trials (420 and 402 episodes). On ToolBench, we cap each episode to 10 steps and run three trials over 450 tasks in total (20 per subset for most splits and 50 for the multi-tool instruction split). For AgentMark-F, we report bits/step and bits/task by summing embedded substrings across steps (with an 8-bit RLNC-coded payload on ToolBench). For the biased red--green baseline ($\gamma=0.5$, $\delta=2.0$), we report Green\% as the detection statistic: $71.9\pm13.6$ (ALFWorld-ID), $71.5\pm15.2$ (ALFWorld-OOD), and $67.0\pm0.9$ (ToolBench), which indicates detectable bias but, as a zero-bit signal, does not support decoding an explicit identifier.

\begin{table}[!b]
\centering
\small
\setlength{\tabcolsep}{6pt}
\renewcommand{\arraystretch}{1.08}
\resizebox{\columnwidth}{!}{
\begin{tabular}{lccc}
\toprule
Scenario & \makecell{No-Induction\\SR (\%)$\uparrow$} & \makecell{AgentMark-F\\SR (\%)$\uparrow$} & \makecell{Baseline\\SR (\%)$\uparrow$} \\
\midrule
\makecell{ALFWorld\\ID} & $83.1 \pm 1.4$ & $89.3 \pm 30.9$ & $89.5 \pm 30.6$ \\
\makecell{ALFWorld\\OOD} & $94.5 \pm 0.9$ & $97.5 \pm 15.6$ & $96.8 \pm 17.7$ \\
ToolBench & $59.6 \pm 8.8$ & $59.7 \pm 4.8$ & $59.9 \pm 5.8$ \\
\bottomrule
\end{tabular}
}
\caption{No-induction calibration on DeepSeek-Chat. The no-induction agent uses a standard ReAct loop without explicit probability elicitation. SR is averaged over three runs on ALFWorld (274 tasks) and ToolBench (120 tasks).}
\label{tab:no_induction}
\end{table}

\subsubsection{No-Induction Calibration}
\label{sec:no_induction}
This supplementary experiment isolates the effect of probability elicitation by comparing AgentMark-F and the elicited baseline against a standard no-induction ReAct loop. The no-induction baseline uses the original greedy ReAct interface without explicit probability elicitation and is repeated three times on ALFWorld (274 tasks) and ToolBench (120 tasks) with DeepSeek-Chat. As shown in Table~\ref{tab:no_induction}, the no-induction baseline is consistently slightly below both Baseline and AgentMark-F, which is consistent with the elicitation prompt encouraging more structured deliberation over candidate behaviors. The key comparison remains Baseline versus AgentMark-F under the same elicitation interface, where the success-rate gap is negligible, providing indirect evidence that the elicited $P_t$ faithfully reflects the underlying planning tendencies.
\FloatBarrier

\subsubsection{Cross-Run Distribution Stability}
\label{sec:jsd_stability}
We additionally measure the stability of elicited behavior distributions across repeated runs using Jensen--Shannon divergence (JSD). On ToolBench, we use DeepSeek-Chat and collect 542 valid task pairs across six subsets and three independent runs. On ALFWorld, we use Gemini-2.0-Flash and collect 140 task trajectories across three runs. Table~\ref{tab:jsd_stability} shows that the first-step distributions are highly consistent across runs ($\mathrm{JSD}\approx 0.07$--$0.08$), while later-step divergence increases as execution histories branch. The overall divergence remains lower on ToolBench than on ALFWorld, which is consistent with the more constrained action spaces in tool-use environments.

\begin{table}[htbp]
\centering
\small
\setlength{\tabcolsep}{6pt}
\renewcommand{\arraystretch}{1.08}
\resizebox{\columnwidth}{!}{
\begin{tabular}{llccc}
\toprule
Dataset & Model & \makecell{First-Step\\JSD} & \makecell{Overall\\JSD} & \makecell{Task\\Pairs} \\
\midrule
ToolBench & DeepSeek-Chat & 0.0701 & 0.1764 & 542 \\
ALFWorld & Gemini-2.0-Flash & 0.0765 & 0.4597 & 140 \\
\bottomrule
\end{tabular}
}
\caption{Cross-run stability of elicited behavior distributions, measured by Jensen--Shannon divergence (JSD) between independently repeated runs.}
\label{tab:jsd_stability}
\end{table}
\begin{table*}[!t]
\centering
\small
\setlength{\tabcolsep}{5pt}
\renewcommand{\arraystretch}{1.05}
\resizebox{\textwidth}{!}{
\begin{tabular}{llccccccc}
\toprule
& & \multicolumn{2}{c}{SR (\%)$\uparrow$} & \multicolumn{2}{c}{Steps} & \multicolumn{2}{c}{Capacity} & \\
\cmidrule(lr){3-4} \cmidrule(lr){5-6} \cmidrule(lr){7-8}
Setting & Task & Base & Ours & Base & Ours & bps & bpt & \makecell{$\Delta$s/step} \\
\midrule
\makecell{ALFWorld\\ID} & Avg. & $70.6 \pm 45.6$ & $67.8 \pm 46.8$ & $25.3 \pm 18.0$ & $28.0 \pm 17.8$ & $2.28 \pm 1.29$ & $80.6 \pm 79.1$ & $-0.18 \pm 1.1$ \\
\makecell{ALFWorld\\OOD} & Avg. & $71.7 \pm 45.1$ & $72.8 \pm 44.6$ & $26.5 \pm 17.3$ & $25.5 \pm 17.5$ & $1.96 \pm 1.1$ & $63.4 \pm 66.6$ & $-0.01 \pm 1.2$ \\
ToolBench & Avg. & $75.6 \pm 8.0$ & $75.8 \pm 4.5$ & $2.66 \pm 0.41$ & $2.74 \pm 0.20$ & $0.78 \pm 0.1$ & $2.1 \pm 0.5$ & $-0.64 \pm 0.5$ \\
\bottomrule
\end{tabular}
}
\caption{Gemini 2.0 Flash on ALFWorld and ToolBench.}
\label{tab:cross_model_results}
\end{table*}
\subsubsection{Cross-Model Results}
\label{sec:cross_model}
We further validate AgentMark on Gemini 2.0 Flash using 420 ALFWorld-ID episodes, 402 ALFWorld-OOD episodes, and results aggregated over six ToolBench subsets, with three runs per setting. As shown in Table~\ref{tab:cross_model_results}, success rates remain close to the corresponding baselines across both ALFWorld and ToolBench, while $\Delta$s/step remains near zero. These results complement Table~\ref{tab:main_results} and confirm that utility preservation and watermark capacity are not specific to DeepSeek-Chat.

\begin{table*}[!t]
\centering
\scriptsize
\renewcommand{\arraystretch}{1.02}
\setlength{\tabcolsep}{4pt}
\resizebox{0.85\textwidth}{!}{
\begin{tabular}{llcccccc}
\toprule
& & \multicolumn{3}{c}{DeepSeek-Chat} & \multicolumn{3}{c}{Gemini 2.0 Flash} \\
\cmidrule(lr){3-5} \cmidrule(lr){6-8}
Setting & Task & \makecell{$\bar c_t$\\(Mean$\pm$Std)} & \makecell{$\overline{H(P_t)}$\\(Mean$\pm$Std)} & Steps & \makecell{$\bar c_t$\\(Mean$\pm$Std)} & \makecell{$\overline{H(P_t)}$\\(Mean$\pm$Std)} & Steps \\
\midrule
\multirow{6}{*}{\makecell{ALFWorld\\ID}} & A1 & $1.78 \pm 1.40$ & $2.42 \pm 1.12$ & 282 & $3.28 \pm 2.05$ & $3.67 \pm 1.66$ & 1246 \\
 & A2 & $1.80 \pm 1.38$ & $2.43 \pm 1.17$ & 479 & $2.89 \pm 2.19$ & $3.33 \pm 1.83$ & 2078 \\
 & A3 & $1.72 \pm 1.30$ & $2.43 \pm 1.05$ & 604 & $2.80 \pm 1.76$ & $3.22 \pm 1.38$ & 2777 \\
 & A4 & $1.36 \pm 1.24$ & $2.25 \pm 1.03$ & 99 & $2.31 \pm 1.80$ & $2.66 \pm 1.50$ & 997 \\
 & A5 & $1.88 \pm 1.32$ & $2.65 \pm 1.11$ & 453 & $2.84 \pm 1.74$ & $3.27 \pm 1.37$ & 1787 \\
 & A6 & $1.82 \pm 1.39$ & $2.67 \pm 1.17$ & 472 & $3.32 \pm 2.14$ & $3.82 \pm 1.67$ & 2254 \\
\midrule
\multirow{6}{*}{\makecell{ALFWorld\\OOD}} & A1 & $1.91 \pm 1.37$ & $2.72 \pm 1.07$ & 479 & $1.69 \pm 1.95$ & $2.40 \pm 1.47$ & 961 \\
 & A2 & $1.67 \pm 1.31$ & $2.42 \pm 1.06$ & 610 & $0.78 \pm 1.49$ & $3.38 \pm 1.49$ & 2823 \\
 & A3 & $1.78 \pm 1.30$ & $2.60 \pm 1.03$ & 502 & $2.20 \pm 1.77$ & $2.81 \pm 1.29$ & 1276 \\
 & A4 & $1.64 \pm 1.27$ & $2.61 \pm 1.05$ & 171 & $3.13 \pm 1.54$ & $3.46 \pm 1.20$ & 1751 \\
 & A5 & $2.24 \pm 1.34$ & $2.93 \pm 1.06$ & 429 & $3.14 \pm 1.78$ & $3.90 \pm 1.12$ & 2247 \\
 & A6 & $1.69 \pm 1.36$ & $2.39 \pm 1.12$ & 306 & $2.06 \pm 1.93$ & $2.57 \pm 1.57$ & 754 \\
\midrule
\multirow{6}{*}{ToolBench} & T1 & $0.95 \pm 1.04$ & $1.42 \pm 0.98$ & 247 & $0.95 \pm 1.15$ & $1.52 \pm 1.06$ & 181 \\
 & T2 & $1.28 \pm 1.13$ & $1.79 \pm 1.02$ & 312 & $0.56 \pm 1.00$ & $1.16 \pm 1.01$ & 163 \\
 & T3 & $0.79 \pm 1.11$ & $1.38 \pm 1.05$ & 224 & $0.61 \pm 0.82$ & $1.14 \pm 0.81$ & 171 \\
 & T4 & $1.47 \pm 1.15$ & $1.97 \pm 1.04$ & 329 & $0.96 \pm 1.14$ & $1.46 \pm 1.06$ & 167 \\
 & T5 & $1.28 \pm 1.15$ & $1.82 \pm 1.06$ & 287 & $0.95 \pm 1.08$ & $1.48 \pm 1.02$ & 176 \\
 & T6 & $1.44 \pm 1.03$ & $1.89 \pm 0.96$ & 416 & $1.10 \pm 1.06$ & $1.64 \pm 0.99$ & 185 \\
\bottomrule
\end{tabular}
}
\caption{Per-task step-level capacity--entropy analysis for DeepSeek-Chat and Gemini 2.0 Flash. $\bar c_t$ and $\overline{H(P_t)}$ are averaged over embedding-active steps only, i.e., steps where the selected bin size exceeds 1 and at least one bit is embedded. Zero-capacity steps are excluded so that the per-step capacity and entropy are computed over the same step population, enabling a direct comparison against the theoretical bound $\bar c_t \le \overline{H(P_t)}$. By contrast, the bps values in Tables~\ref{tab:main_results} and~\ref{tab:cross_model_results} divide total embedded bits by all decision steps, including zero-capacity ones, yielding lower averages.}
\label{tab:capacity_entropy_panels}
\end{table*}

\subsubsection{Capacity--Entropy and Variance Statistics}
\label{sec:capacity_entropy}
To validate the theoretical capacity bound, we restrict the analysis to embedding-active steps, namely steps where the differential recombination produces a bin of size $n \ge 2$, so that at least one bit is embedded. Steps with a singleton bin ($n = 1$) contribute zero capacity and carry an undefined per-step entropy-to-capacity ratio; including them would conflate the capacity--entropy relationship with the proportion of trivial steps. Moreover, restricting to embedding-active steps ensures that the entropy $\overline{H(P_t)}$ is computed over the same decision points where the watermarking algorithm actually operates, yielding a meaningful upper-bound test. By conditioning on the same step set for both $\bar c_t$ and $\overline{H(P_t)}$, we obtain a paired comparison that directly tests whether $\bar c_t \le \overline{H(P_t)}$ holds step by step. Note that the benchmark-level bps reported in Table~\ref{tab:main_results} averages total embedded bits over all decision steps, including zero-capacity ones, and is therefore systematically lower than $\bar c_t$. Table~\ref{tab:capacity_entropy_panels} confirms this relationship across all 36 task--model combinations and further confirms that higher-entropy settings yield higher empirical capacity. For example, ALFWorld-OOD A5 exhibits substantially larger entropy and capacity than ToolBench T3, while the larger Gemini capacities align with its broader planning distributions under agent-oriented fine-tuning.

\paragraph{Prompt Examples}
\label{appendix:prompt-examples}

\begin{tcolorbox}[breakable, title=ALFWorld Agent Prompt]
\raggedright \small

\textbf{System Prompt} \par \smallskip
{\ttfamily You are an expert household task agent. Analyze the current situation and assign probabilities to each action.} \par \bigskip

\textbf{Few-Shot Examples} \par \smallskip
\textit{Context:} {\ttfamily \{few\_shot\_examples\_based\_on\_task\_type\}} \par \bigskip

\textbf{Success Action Blueprints} \par \smallskip

\textit{pick\_and\_place\_simple:} \\
{\ttfamily 
\quad 1. Search locations like cabinet/drawer/countertop for the target object. \\
\quad 2. take <target> \\
\quad 3. go to <destination> \\
\quad 4. move <target> to <destination> 
} \par \smallskip

\textit{pick\_clean\_then\_place:} \\
{\ttfamily 
\quad 1. Find and take the target object. \\
\quad 2. go to sinkbasin 1 \\
\quad 3. clean <target> with sinkbasin 1 \\
\quad 4. go to <destination> \\
\quad 5. move <target> to <destination>
} \par \smallskip

\textit{pick\_heat\_then\_place:} \\
{\ttfamily 
\quad 1. Find and take the target object. \\
\quad 2. go to microwave 1 \\
\quad 3. heat <target> with microwave 1 \\
\quad 4. go to <destination> \\
\quad 5. move <target> to <destination>
} \par \smallskip

\textit{pick\_cool\_then\_place:} \\
{\ttfamily 
\quad 1. Find and take the target object. \\
\quad 2. go to fridge 1 \\
\quad 3. cool <target> with fridge 1 \\
\quad 4. go to <destination> \\
\quad 5. move <target> to <destination>
} \par \smallskip

\textit{pick\_two\_obj\_and\_place:} \\
{\ttfamily 
\quad 1. take <target1> $\rightarrow$ go to <destination> $\rightarrow$ move <target1> to <destination> \\
\quad 2. Return to origin $\rightarrow$ take <target2> $\rightarrow$ go to <destination> $\rightarrow$ move <target2> to <destination>
} \par \smallskip

\textit{look\_at\_obj\_in\_light:} \\
{\ttfamily 
\quad 1. Find and take the target object (e.g., bowl/cd). \\
\quad 2. go to the observed location of the desklamp. \\
\quad 3. use desklamp (no need to put the object down). \\
\quad 4. examine <target>
} \par \bigskip

\textbf{Current Situation} \par \smallskip
\textit{Recent History:} {\ttfamily \{interaction\_history\_last\_5\_steps\}} \par \smallskip
\textit{Inventory:} {\ttfamily \{inventory\_status\_and\_checks\}} \\
\textit{Task Goal:} {\ttfamily \{task\_description\}} \\
\textit{Observation:} {\ttfamily \{observation\}} \par \smallskip
\textit{Available Actions:} {\ttfamily \{admissible\_commands\_json\_list\}} \par \bigskip

\textbf{Response Format} \par \smallskip
\textit{Thinking:} {\ttfamily Write a 'Thinking: ...' section to analyze the situation.} \\
\textit{Output:} {\ttfamily Output the JSON probability object (sum to 1.0).}
\end{tcolorbox}

\begin{tcolorbox}[breakable, title=Toolbench Agent Prompt]
\raggedright \small 

\textbf{System Prompt} \par \smallskip
{\ttfamily You are a tool-use agent. First write a short Thought, then output a JSON block only with keys: "action\_weights" (probabilities for EVERY candidate tool, sum to 1) and "action\_args" (arguments for EVERY candidate tool; if Finish prob is low, keep its args null). Do not output prose outside the Thought + JSON. Use only the provided tool names. If a tool returns an error or no data, please inform the user truthfully and do not make up information.} \par \bigskip

\textbf{User Prompt} \par \smallskip
\textit{User Query:} {\ttfamily \{query\}} \\
\textit{Tools:} {\ttfamily \{tools\_block\}} \par \smallskip

{\ttfamily Respond with a Thought line, then a JSON block. Include probabilities and arguments for EVERY tool.} \par \smallskip

\textbf{Example:} \\
\textit{Thought:} {\ttfamily I have checked the API health and it is operational. Now I need to retrieve the project list to answer the user's request. I should not finish yet because I haven't got the list.} \par \smallskip

{\ttfamily 
\{ \\
\quad "action\_weights": \{ \\
\qquad "Tool\_A": 0.1, "Tool\_B": 0.1 ... \\
\quad \}, \\
\quad "action\_args": \{ \\
\qquad "Tool\_A": \{"q": "good\_args\_A"\}, \\
\qquad "Tool\_B": \{"q": "good\_args\_B"\}, \\
\qquad "Finish": \{"final\_answer": null\} \\
\quad \} \\
\} 
} \par \bigskip

\textit{Instructions:} \\
{\ttfamily Ensure every candidate tool (including Finish) appears in action\_weights and action\_args. If Finish prob is low, keep its args empty/null; when you choose Finish, put the final\_answer there.}

\end{tcolorbox}

\subsection{Details in OASIS}
\label{appendix:utility-capacity-oasis}

\subsubsection{Agent simulation setup}
\label{appendix:oasis-setup}
We evaluate AgentMark-F in OASIS~\cite{yang2024oasis} on two platforms, a Twitter-like environment and a Reddit-like environment.
For each platform, we collect $100$ agent trajectories in total.
Each run simulates $10$ agents for $10$ environment steps, including $5$ agents with AgentMark-F enabled and $5$ no-watermark control agents, and we repeat runs with different random seeds to reach the target trajectory count.
All agents use the same base LLM (\texttt{deepseek-chat}) with a maximum context length of 16000 tokens, and the same scenario scripts and initialization rules across watermark and control groups.
At each step, we log the elicited behavior distribution $P_t$ over a fixed behavior list and the selected planning behavior, which are sufficient for AgentMark decoding and for computing Watermark Detection Rate.

The admissible behavior list is platform-specific.
For the Twitter-like platform, the behavior set includes \texttt{create\_post}, \texttt{like\_post}, \texttt{repost}, \texttt{follow}, \texttt{quote\_post}, and \texttt{do\_nothing}.
For the Reddit-like platform, the behavior set includes posting and commenting actions, vote actions, search and refresh actions, and social actions such as \texttt{follow} and \texttt{mute}.
To simulate realistic social dynamics, we use a random follow graph in which each agent follows $3$--$5$ other agents.

We define the step context used for reproducible randomness using a short recent interaction trace (the most recent $3$ behaviors), which the verifier can reconstruct from the same logged records.

\subsubsection{Prompt examples}
\label{appendix:oasis-prompts}

\begin{tcolorbox}[breakable, title=Twitter Agent Prompt]
\raggedright \small

\textbf{System Message} \par \smallskip

\textit{Objective:} \\
{\ttfamily You're a Twitter user, and I'll present you with some posts. After you see the posts, choose some actions from the following functions.} \par \smallskip

\textit{Self-Description:} \\
{\ttfamily Your actions should be consistent with your self-description and personality. Your name is \{name\}. Your profile: \{profile\}.} \par \smallskip

\textit{Language Preference:} \\
{\ttfamily Please output all your comments, posts, and reasoning in Chinese.} \par \smallskip

\textit{Response Method:} \\
{\ttfamily Please perform actions by tool calling.} \par \bigskip

\textbf{User Message} \par \smallskip

\textit{Context Observation:} \\
{\ttfamily You are observing a social media environment: \{env\_prompt\}} \par \smallskip

\textit{Task Instruction:} \\
{\ttfamily Based on this observation and your profile, estimate the probability of performing each action. Return ONLY a JSON object with probabilities (must sum to 1.0).} \par \smallskip

\textit{Available Actions:} \\
{\ttfamily create\_post, like\_post, repost, follow, do\_nothing, quote\_post} \par \smallskip

\textit{Output Format:} \\
{\ttfamily \{"action\_name": probability, ...\}} \par \smallskip

\textit{Example:} \\
{\ttfamily
\{ \\
\quad "like\_post": 0.3, \\
\quad "create\_post": 0.25, \\
\quad "follow": 0.2, \\
\quad "do\_nothing": 0.25 \\
\}
}

\end{tcolorbox}

\begin{tcolorbox}[breakable, title=Reddit Agent Prompt]
\raggedright \small

\textbf{System Message} \par \smallskip

\textit{Objective:} \\
{\ttfamily You're a Reddit user, and I'll present you with some posts. After you see the posts, choose some actions from the following functions.} \par \smallskip

\textit{Self-Description:} \\
{\ttfamily Your actions should be consistent with your self-description and personality. Your name is \{name\}. Your profile: \{profile\}.} \par \smallskip

\textit{Language Preference:} \\
{\ttfamily Please output all your comments, posts, and reasoning in Chinese.} \par \smallskip

\textit{Response Method:} \\
{\ttfamily Please perform actions by tool calling.} \par \bigskip

\textbf{User Message} \par \smallskip

\textit{Context Observation:} \\
{\ttfamily You are observing a social media environment: \{env\_prompt\}} \par \smallskip

\textit{Task Instruction:} \\
{\ttfamily Based on this observation and your profile, estimate the probability of performing each action. Return ONLY a JSON object with probabilities (must sum to 1.0).} \par \smallskip

\textit{Available Actions:} \\
{\ttfamily \{actions\_str\}} \par \smallskip

\textit{Output Format:} \\
{\ttfamily \{"action\_name": probability, ...\}} \par \smallskip

\textit{Example:} \\
{\ttfamily
\{ \\
\quad "like\_post": 0.3, \\
\quad "create\_comment": 0.25, \\
\quad "follow": 0.2, \\
\quad "refresh": 0.25 \\
\}
}

\end{tcolorbox}

\section{Robustness Experiment}

\subsection{Details of False Positives and Key Forgery}
\label{appendix:fpr}

\paragraph{Goal.}
We quantify how often an invalid record passes verification under two settings: (i) unwatermarked logs, modeled by random GF(2) outputs, and (ii) wrong-key decoding, modeled by generating the RLNC coefficients with an incorrect key.

\paragraph{Data and coefficient construction.}
We aggregate 281 predicted trajectories from ToolBench (T1--T3) and extract 159 unique action indices.
These indices are used to construct the GF(2) coefficient matrix $A$ so that the sparsity pattern and coefficient distribution reflect the deployment setting rather than an idealized fully random matrix.

\paragraph{Verification rule.}
Let the payload length be $N=128$ and let the total number of received packets be $m=N+k$ with overhead $k\in[0,16]$.
For each trial, we form $A\in\mathbb{F}_2^{m\times N}$ and a vector $y\in\mathbb{F}_2^{m}$.
We accept if and only if the system $Ax=y$ is consistent over $\mathbb{F}_2$, i.e., $\mathrm{rank}(A)=\mathrm{rank}([A\mid y])$.

\paragraph{Simulation protocol.}
For each $k$, we run 1000 independent Monte Carlo trials.
In the unwatermarked setting, $y$ is sampled uniformly at random.
In the wrong-key setting, $A$ (and thus the induced equations) is generated using an incorrect key.
We estimate FPR as the fraction of trials that pass the consistency check and report 95\% confidence intervals using $\pm 1.96\times\mathrm{SEM}$.

\paragraph{Results.}
Table~\ref{tab:fpr_table} and Figure~\ref{fig:fpr} show that both unwatermarked and wrong-key FPRs decrease rapidly with $k$ and closely follow an exponential trend.
Concretely, FPR drops from $60.6\%$ at $k=0$ to below $1\%$ at $k=8$ (0.40\% unwatermarked; 0.50\% wrong key), and we observe no false positives for $k\ge 14$ in 1000 trials.

\begin{table}[t]
\centering
\small
\setlength{\tabcolsep}{6pt}
\renewcommand{\arraystretch}{1.1}
\begin{tabular}{ccc}
\toprule
Overhead $k$ & Unwatermarked & Wrong Key \\
\midrule
0  & 60.60\% & 60.60\% \\
2  & 22.70\% & 23.30\% \\
4  & 7.00\%  & 5.90\%  \\
6  & 1.10\%  & 1.20\%  \\
8  & 0.40\%  & 0.50\%  \\
$\ge 14$ & 0.00\% & 0.00\% \\
\bottomrule
\end{tabular}
\caption{False-positive rates under the consistency-check verifier for unwatermarked logs and wrong-key decoding.}
\label{tab:fpr_table}
\end{table}

\subsection{Details of Robustness to Step Erasure and Truncation}
\label{appendix:robustness}

\paragraph{Goal and setting.}
We study robustness under missing planning-time records by simulating step erasure on logged ToolBench trajectories.
We focus on recovering an $8$-bit deployment-level ID payload ($k=8$) from incomplete logs, and compare RLNC-based coding with a repetition-code baseline under both single-episode and global decoding.

\paragraph{Step-erasure model.}
For each logged trajectory, we randomly and independently drop each decision step with probability $p\in[0,1]$.
When a step is erased, all packets/equations emitted at that step are removed together, yielding an observed subsequence consistent with log loss or truncation effects.

\paragraph{AgentMark-F with RLNC coding.}
Under RLNC, each surviving step contributes a variable number of linear measurements (packets) depending on the per-step capacity.
The verifier stacks all received measurements into a linear system over $\mathbb{F}_2$ and declares success if the recovered system has full rank (equivalently, if the decoded payload is uniquely determined).
This directly matches our design goal that decoding depends primarily on the total received packet budget rather than fixed block alignment.

\paragraph{Repetition-code baseline.}
As a baseline, we repeatedly transmit the $k$ payload bits across steps.
We use a strict blind-decoding setting with no auxiliary indices/sequence numbers, and the decoder succeeds only if it can recover an intact length-$k$ payload segment from the surviving packet stream (details below).
This baseline is included to highlight the advantage of rateless linear measurements under step erasures.

\paragraph{Single-episode versus global decoding.}
We consider two verification regimes.
\emph{Single-episode decoding} attempts to decode the payload from each trajectory independently, which provides a conservative lower bound when logs are fragmented.
\emph{Global decoding} aggregates surviving packets across multiple trajectories into a single stream and decodes once, which emulates deployment-oriented auditing where a verifier pools logs over a time window and therefore benefits from cross-trajectory redundancy.

\paragraph{Task filtering for fair comparison.}
To ensure a fair robustness comparison, we filter trajectories so that both methods achieve $100\%$ decoding success at $p=0$.
For RLNC, we select $25$ sufficiently long trajectories that can support full recovery without erasure.
For the repetition baseline, we filter to $19$ long trajectories (e.g., requiring at least $k$ packets/steps) so that blind decoding is feasible at $p=0$.

\paragraph{Evaluation protocol and confidence intervals.}
For each erasure probability $p$, we repeat the random erasure process $30$ times per trajectory and record whether decoding succeeds.
This yields $25\times 30=750$ trials for RLNC single-episode and $19\times 30=570$ trials for repetition single-episode at each $p$.
For global decoding, we repeat global aggregation under $30$ independent erasure realizations.
We report the mean decode success rate and a $95\%$ confidence interval computed using the Wilson score interval for a Bernoulli proportion.
We sweep $p$ over $[0,1]$ with a finer grid near the phase-transition region to visualize the sharp robustness boundary.

\subsection{Semantic Rewriting}
\label{appendix:semantic}

\paragraph{Details}
We report additional details for the semantic-preserving observation rewriting stress test used in Section~\ref{sec:exp-utility-capacity}.
The goal is to quantify how sensitive planning-time behavior selection and behavioral watermark decoding are when the \emph{semantics} of the environment observation is preserved but its surface form is rewritten, which can arise when raw logs are unavailable and only reconstructed records are retained.

We run the test on ALFWorld-OOD (134 tasks; 2326 steps) using \texttt{deepseek-chat} with temperature $1.0$.
At each step, we rewrite the \texttt{[CURRENT SITUATION]} portion of the observation with an LLM under the constraint that no objects or state information are added or removed, and only wording and syntax change.
We evaluate under a teacher-forcing protocol that keeps the same original history/context at every step, so that the measured differences isolate the effect of observation rewriting rather than compounding trajectory drift.

\paragraph{Prompt example}
\begin{tcolorbox}[breakable, title=Semantic Rewriting Prompt]
\raggedright \small

\textbf{System Prompt} \par \smallskip
{\ttfamily You are a helpful assistant. Rewrite the following text to convey exactly the same information but using different words and sentence structures. Do not add or remove any facts. Keep it concise.} \par \bigskip

\textbf{User Prompt} \par \smallskip
\textit{Input Text:} {\ttfamily [The original observation context goes here]}
\end{tcolorbox}

\subsection{Adaptive Threat Models}
\label{sec:adaptive_adversary}
The experiments in the main paper and appendix focus on random erasure, truncation, and semantic-preserving rewriting, where the attacker does not adapt to the watermarking algorithm itself. A stronger threat model is a fully adaptive adversary who knows the watermarking scheme and deliberately edits step context or trajectories to disrupt synchronization and reduce the number of usable measurements. We view such adaptive attacks as an important future direction, especially for settings where the adversary can observe verification failures and iteratively refine the perturbation strategy. Our current results under semantic rewriting provide a first stress test in this direction, but they do not exhaust the adaptive threat space.

\section{Compatibility Experiment Details}
\label{appendix:compat}

\paragraph{Environment and trajectories.}
We conduct the compatibility experiment on ToolBench~\cite{qin2023toolllm} using \texttt{Llama-3.2-3B-Instruct} as the base model.
Each episode is capped at a maximum of 25 decision steps, and the agent may terminate earlier upon completion.
We generate 149 trajectories in total, and log the planning-time behavior choices (for AgentMark-F decoding) together with the final action-level textual outputs (for content-watermark detection).

\paragraph{Watermarking configuration.}
We enable AgentMark-F as a behavior-level watermark by embedding multi-bit provenance into planning behaviors via distribution-preserving conditional sampling, while leaving the downstream action execution unchanged.
Independently, we apply SynthID-Text~\cite{dathathri2024scalable} as an action-level (text) watermark on the final surface-form outputs produced by the agent.
For SynthID-Text, we follow the open-source MarkLLM toolkit configuration~\cite{pan2024markllm} with distortionary watermark mode, $n$-gram length 5, 4 leaves, and 30 keys.
For text generation in the watermark setting, we use temperature 1.0 as configured in MarkLLM.

\paragraph{Detection and decoding.}
We evaluate SynthID-Text detectability on the collected ToolBench outputs using the corresponding MarkLLM detector under the same configuration. The watermark detection rate is $96.64 \pm 0.06\%$.
In parallel, we verify that AgentMark-F remains decodable from the logged planning-time behaviors and their associated behavior distributions, which demonstrates that enabling the action-level content watermark does not disrupt behavior-level provenance recovery.
Overall, this setup tests composability in a realistic agent pipeline where behavior watermarking and content watermarking operate on different decision layers and are evaluated through their respective interfaces (trajectory logs versus final text outputs).

\end{document}